# (SARS-CoV-2) COVID 19: Genomic surveillance and evaluation of the impact on the population speaker of indigenous language in Mexico

Carlos Medel-Ramírez
Universidad Veracruzana / Instituto de Investigaciones y Estudios Superiores Económicos y Sociales / Observes
ORC ID: 0000-0002-5641-6270
cmedel@uv.mx

Hilario Medel-López
Universidad Veracruzana / Instituto de Antropología
ORC ID: 0000-0002-0072-8654
hmedel@uv.mx

Jennifer Lara Mérida
Universidad de Xalapa / Facultad de Derecho
ORC ID: 0000-0003-2121-5652
jenniferlaram@gmail.com

## Resumen

La importancia del documento de trabajo radica en que permite el análisis de la información y de casos asociados con el (SARS-CoV-2) COVID-19, a partir de la información diaria generada por el Gobierno de México a través de la Secretaría de Salud, responsable del Sistema de Vigilancia Epidemiológica de Enfermedades Respiratoria Viral (SVEERV). La información en el SVEERV se difunde como datos abiertos (open data), y el nivel de información se muestra a nivel municipal, estatal y nacional. Por otra parte, el segumiento de la vigilancia genómica del (SARS-CoV-2) COVID-19, mediante la identificación variantes y mutaciones se registra en la base datos del Sistema de Información de la Global Initiative on Sharing All Influenza Data (GISAID) con sede de Alemania. Estas dos fuentes del información SVEERV y GISAID aportan la información para el análisis del impacto del el (SARS-CoV-2) COVID-19 en la población en México. En la primera fuente de datos se identifica información, a nivel nacional, de pacientes según edad, sexo, comorbilidades y condición de presencia (SARS-CoV-2) COVID-19, entre otras características.

El análisis de datos se realiza mediante el diseño de un algoritmo aplicando técnicas y metodología de minería de datos, para estimar la tasa de letalidad, indice de positividad e identificar una tipología atendiendo a la gravedad de la infección identificada en pacientes quienes presentan un resultado positivo para (SARS-CoV-2) COVID-19. De la segunda fuente de datos, se obtiene información a nivel mundial de las nuevas variantes y mutaciones del (SARS-CoV-2) COVID-19, aportando información valiosa para la vigilancia genómica oportuna. El presente estudio analiza el impacto del (SARS-CoV-2) COVID-19 en la población hablante de lengua indígena permite proporcionar información, de forma rápida y oportuna para apoyar el diseño de politica pública en materia de salud.

## Resume

The importance of the working document is that it allows the analysis of information and cases associated with (SARS-CoV-2) COVID-19, based on the daily information generated by the Government of Mexico through the Secretariat of Health, responsible for the Epidemiological Surveillance System for Viral Respiratory Diseases (SVEERV). The information in the SVEERV is disseminated as open data, and the level of information is displayed at the municipal, state and national levels. On the other hand, the monitoring of the genomic surveillance of (SARS-CoV-2) COVID-19, through the identification of variants and mutations, is registered in the database of the Information System of the Global Initiative on Sharing All Influenza Data (GISAID) based in Germany. These two sources of information SVEERV and GISAID provide the information for the analysis of the impact of (SARS-CoV-2) COVID-19 on the population in Mexico. The first data source identifies information, at the national level, on patients according to age, sex, comorbidities and COVID-19 presence (SARS-CoV-2), among other characteristics. The data analysis is carried out by means of the design of an algorithm applying data mining techniques and methodology, to estimate the case fatality rate, positivity index and identify a typology according to the severity of the infection identified in patients who present a positive result. for (SARS-CoV-2) COVID-19. From the second data source, information is obtained worldwide on the new variants and mutations of COVID-19 (SARS-CoV-2), providing valuable information for timely genomic surveillance. This study analyzes the impact of (SARS-CoV-2) COVID-19 on the indigenous language-speaking population, it allows us to provide information, quickly and in a timely manner, to support the design of public policy on health.

## Introducción

The use of data mining models has made it possible to analyze information on the presence of (SARS-CoV-2) COVID 19 worldwide and particularly in Mexico. The analysis of the social problems in public health caused by the (SARS-CoV-2) COVID 19 directs the actions of the scientific community to identify the possible solution, in the face of the crisis scenario that occurs in public health systems. The vertiginous advance of the investigations on the (SARS-CoV-2) COVID 19, has allowed a change in the paradigm of the disclosure and publication of the results of scientific research, by virtue of which the urgent need of a greater speed is weighed in the communication of research results, a situation that has made it possible to open the frontiers of knowledge and create opportunities for scientific development, since the purpose is to attend to the health emergency, whose dimension is global. The presence of (SARS-CoV-2) COVID 19 in the territory of Mexico[1] It invites us to the following reflections: how has (SARS-CoV-2) COVID 19 impacted the indigenous language-speaking population in Mexico? And what genomic variants have been identified in Mexico from April 2020 to September 2021?

This document analyzes the information on the presence of (SARS-CoV-2) COVID 19 in Mexico, using data mining techniques to design an algorithm [2] that allows evaluating the impact of (SARS-CoV-2) COVID 19 on the indigenous language-speaking population in Mexico, analyzing the fatality rate, positive indication, level of infection severity as well as the identification of variants genomic diseases present in the country, in order to identify the degree of presence of those that show a higher level of contagion.

[1] On February 27, 2020, the first case of (SARS-CoV-2) COVID 19 was registered in Mexico. According to the registry of (SARS-CoV-2) COVID 19, it refers to a patient of Mexican nationality from Italy, who presented mild symptoms. See. The Economist. (2020). "Ministry of Health confirms the first case of coronavirus in Mexico." Section: Politics. February 28, 2020. Retrieved from: https://www.eleconomista.com.mx/politica/Secretaria-de-Salud-confirma-el-primer-caso-de-coronavirus-en-Mexico-20200228-0061.html

[2] This article is based on the proposal presented by Medel-Ramírez, Carlos & Medel-López, Hilario, 2020. "Data Article. Data mining for the study of the Epidemic (SARS-CoV-2) COVID-19: Algorithm for the identification of patients (SARS-CoV-2) COVID 19 in Mexico, "MPRA Paper 100888, University Library of Munich, Germany. This document has been cited in the book by Marie-Odile Safon and Véronique Suhard (2020) “Covid-19 Éléments de littérature scientifique Bibliographie thématique Septembre 2020. Center de documentation del L'Institut de recherche et documentation en économie de la santé (Irdes ), France. " ISBN 978-2-87812-526-9, being considered as a contribution to the study of (SARS-CoV-2) COVID-19 and integrated into the section of international studies related to epidemiological aspects and models of spread of the infection. (See page 54 of the cited document). Retrieved from: https://www.irdes.fr/documentation/syntheses/covid-19-premiers-elements-de-litterature-scientifique.pdf

The analysis of the information has a cut-off date of September 3, 2021, with data from two sources of information, national and international. The data source at the national level comes from the Epidemiological Surveillance System for Viral Respiratory Diseases (SVEERV); while the international source emanates from the Information System of the Global Initiative on Sharing All Influenza Data (GISAID) based in Germany. From the national source of information, it corresponds to the daily record carried out by the Government of Mexico through the Ministry of Health, responsible for the Epidemiological Surveillance System for Viral Respiratory Diseases (SVEERV).[3] The information from SVEERV is disseminated as open data and the information is displayed at the municipal, state and national levels. The monitoring of genomic surveillance of (SARS-CoV-2) COVID-19 is carried out by identifying variants and mutations registered in the database of the Information System of the Global Initiative on Sharing All Influenza Data[4] (GISAID) based in Germany, with information provided by national health authorities in various countries, with a level of periodicity and quality in the identification of genomic variants identified worldwide.

The first data source contains information, at the national, state and municipal level of patients according to age, sex, comorbidities and COVID-19 presence condition (SARS-CoV-2), among other characteristics. The data analysis is carried out through the design of an algorithm applying data mining techniques and methodology, to estimate the case fatality rate, the positivity index and identify a typology according to the severity of the infection identified in patients who present a result positive for (SARS-CoV-2) COVID-19. While from the second data source, information is obtained worldwide on the new variants and mutations of the (SARS-CoV-2) COVID-19, providing valuable information for timely genomic surveillance.[5]

[3] See. Ministry of Health (2021). Information regarding COVID-19 cases in Mexico. Retrieved from: https://datos.gob.mx/busca/dataset/informacion-referente-a-micos-covid-19-en-mexico

[4] See. Global Initiative on Sharing All Influenza Data (GISAID) (2021). Genomic epidemiology of hCoV-19. Recovered from: https://www.gisaid.org/

The importance of this research is its relevance in a line within the National Research and Incidence Projects (Pronaii) and (PRONACES) 2021, which is part of the thematic Axis "Health", on the topic: "Genomic surveillance of SARS -CoV-2 ", corresponding to the subtopic:" SARS-CoV-2 metadata analysis and genomic sequencing ".[6] This study analyzes the impact of (SARS-CoV-2) COVID-19 on the indigenous language-speaking population, allowing information to be provided, quickly and in a timely manner, to support the design of public health policy.

**(SARS-CoV-2) COVID 19:** An emerging coronavirus

Although, since 1968 there has been a record of coronaviruses, the current one (SARS-CoV-2) COVID-19[7] exhibits an exponential contagion level, adaptability and rapid mutation[8]. (Musarrat Abbas K., et al., 2020: 2) point out that (SARS-CoV-2)

---

[5] See. World Health Organization (2021). SARS-CoV-2 genomic sequencing for public health goals. Retrieved from: https://www.who.int/publications/i/item/WHO-2019-nCoV-genomic_sequencing-2021.1

[6] See. National Council for Science and Technology. (2021). "National Research and Incidence Projects (Pronaii)". Recovered from: https://conacyt.mx/pronaces/pronaces-salud/

[7] The term coronavirus was coined in 1968, and is described in the article by Tyrrel, D. A. J., J. D. Almedia, D. M. Berry, C. H. Cunningham, D. Hamre, M. S. Hofstad, L. Malluci, and K. McIntosh. (1968). Coronavirus Nature 220: 650. Retrieved from: https://www.ncbi.nlm.nih.gov/pmc/articles/PMC7182643/

[8] In a report dated November 11, 2020, it is noted that in the production of vaccines to face the (SARS-CoV-2) COVID 19, the following developments take place: 1) AstraZeneca Swedish multinational vaccine developed by the University of Oxford, with an efficiency of 90% in phase III; 2) Sinopharm vaccine from Chinese pharmaceuticals with trials in China, Brazil and Bahrain, in the final testing phase; 3) Pfizer and BioNTech German vaccine with 90% efficiency in phase III; 4) Modern American vaccine with an efficiency of 94.5% in the last phase of clinical trials; 5) Moscow Institute of Epidemiology Russian vaccine with an efficiency of 95% and in final production; and 6) CureVac German experimental vaccine in final testing process. See. Reuters. (2020). What are and in what phase are the possible vaccines against COVID 19. Retrieved from: https://www.eitb.eus/es/noticias/sociedad/detalle/7622841/listado-posibles-vacunas-covid19 -November-2020 /

COVID 19 belongs to the order of Nidovirales and family Coronaviridae [9], from which two subfamilies emerge: a) Coronavirus and b) Torovirus. Within the Coronavirus subfamily, the genes identified correspond to the following: a) VOC Alpha, b) VOC Beta, c) VOC Delta Cov, d) VOC Gamma, e) VOC Delta , f) VOI Eta, g) VOI Iota, h) VOI Kappa, i) VOI Lambda y j) VOI Mu. (See Chart 1, below).

Graph 1 Coronavirus Classification

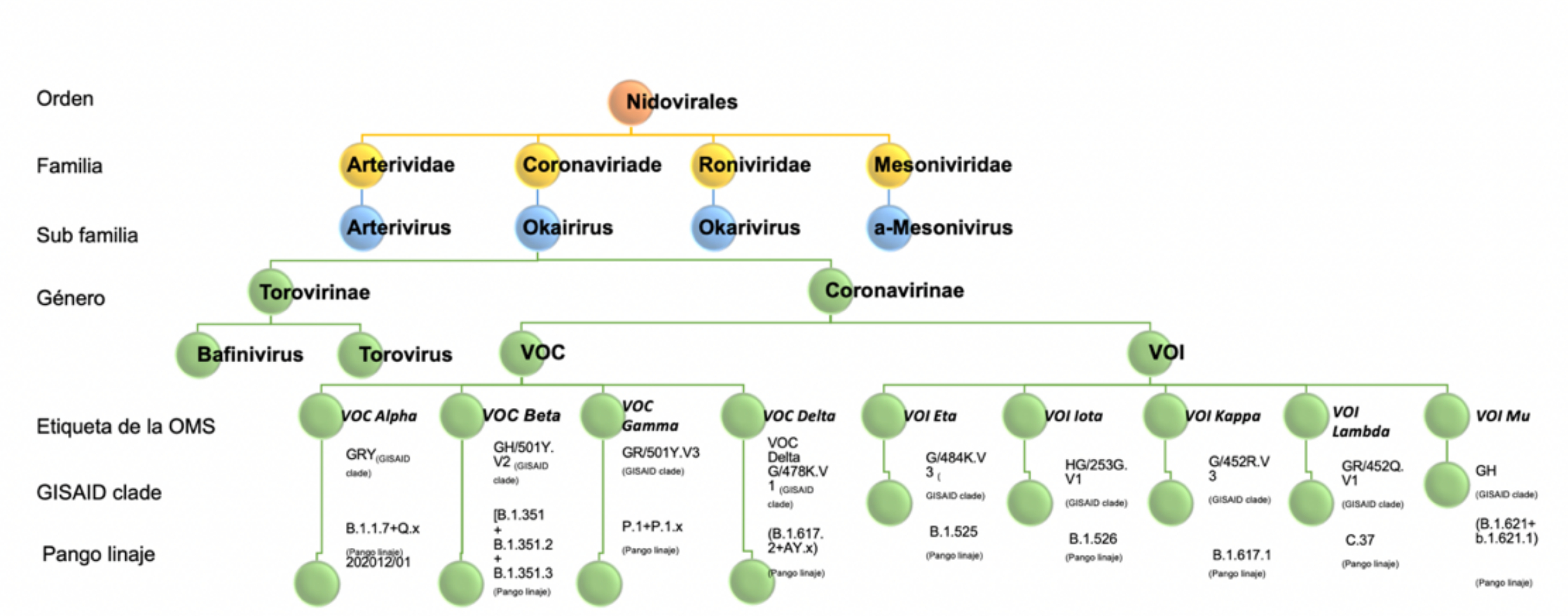

Source: Self made.
Adapted from that proposed in "Junejo Y., et al. (2020). Novel SARS-CoV-2 / COVID 19: Origin, pathogenesis, genes and genetic variations, immune responses and phylogenetic analysis, Gene Reports. Vol. 20, 2020, 100752, p. 2 ", and is updated with information from the" Institute of Social and Preventive Medicine University of Bern, Switzerland & SIB Swiss Institute of Bioinformatics, Switzerland (2021) "and the" Global Initiative on Sharing All Influenza Data (GISAID) (2021). Genomic epidemiology of hCoV-19 ", with a cut-off date of September 3, 2021. Retrieved from: https://www.epicov.org/epi3/frontend#631ee5

At present, new genomic variants of the virus have been registered, identified in priority variants [10] (VOC) and variants of interest [11] (VOI), being the following:

---

[9] The Coronaviridae name is due to the fact that the virus has a kind of crown-shaped spikes. See. 10. Weiner L.P. (1987) Coronaviruses: A Historical Perspective. In: Lai M.M.C., Stohlman S.A. (eds) Coronaviruses. Advances in Experimental Medicine and Biology, vol 218. Springer, Boston, MA. https://doi.org/10.1007/978-1-4684-1280-2_1, p. 2

[10] According to the classification of the Ministry of Health of Argentina, the priority variant identified as (VOC): a) is identified by an increase in transmission and a greater number of cases in a region, b) is identified by an increase in the transmission of contagion or c) is identified by reducing the capacity of control measures, diagnostics or vaccines See. Argentine Ministry of Health (2021). COVID 19 Current situation of the new SARS-CoV-2 variants. Technical report. July 2021. p. 2 Recovered from: https://www.argentina.gob.ar/sites/default/files/2021/07/vigilancia-genomica-se26.pdf

[11] Variants of interest (VOI): It is the identification of a variant according to its phenotypic identification which presents mutations that lead to amino acid changes associated with established phenotypic changes. See. Argentine Ministry of Health (2021). COVID 19 Current situation of the new SARS-CoV-2 variants. Technical report. Op.cit.

1. *[VOC], [**Alpha** (WHO label)],[**GRY**(GISAID clade)], [* **B.1.1.7**+Q.x (Pango linaje)], *202012/01* first detected in Great Britain, December 18, 2020.[12]
2. *[VOC], [**Beta** (WHO label)], [**GH/501Y.V2** (GISAID clade)], [***B.1.351 + B.1.351.2 + B.1.351.3** (Pango linaje) ], first detected in South Africa, December 18, 2020.[13]
3. *[VOC], [**Gamma** (WHO label) ], [**GR/501Y.V3** (GISAID clade) ],* [ **P.1+P.1.x** (Pango linaje)], first detected in Brazil and Japan on January 11, 2021[14]
4. *[VOC], [**Delta** (WHO label) ], [**G/478K.V1** (GISAID clade) ],*[**B.1.617.2+AY.x** (Pango linaje)], first detected in India on May 11, 2021.[15]
5. *[VOI], [**Eta** (WHO label) ], [**G/484K.V3** (GISAID clade)], [ **B.1.525*** (Pango linaje) ], first detected in the UK and Nigeria.[16]
6. *[VOI], [**Iota** (WHO label) ], [ **HG/253G.V1** (GISAID clade) ], [**B.1.526*** (Pango linaje) ], first detected in the United States of America / New York.[17]
7. *[VOI], [**Kappa** (WHO label) ], [**G/452R.V3** (GISAID clade) ], [**B.1.617.1*** (Pango linaje) ], India first detected.[18]
8. *[VOI], [**Lambda** (WHO label) ], [**GR/452Q.V1** (GISAID clade) ], [**C.37*** (Pango linaje) ], first detected in Peru.[19]
9. *[VOI], [**Mu** (WHO label) ], [**GH**(GISAID clade) ], [**B.1.121+b.1.21.1*** (Pango linaje) ], detected for the first time in Colombia.[20]

In Mexico, the genomic variants present as of September 3, 2021 are: VOC Alpha, VOC Beta, VOC Gamma, VOC Delta, VOI Eta, VOI Iota y VOI Lambda. See Graph 2 below, which shows the variants identified in the period from March 2020 to September 2021.

[12]See. Centers for Disease Control and Prevention. (2021). "Emerging SARS-CoV-2 Variants". January 15 2021. Retrieved from: https://www.cdc.gov/coronavirus/2019-ncov/more/science-and-research/scientific-brief-emerging-variants.html

[13] See. Centers for Disease Control and Prevention. (2021). Op. cit

[14] See. Centers for Disease Control and Prevention. (2021). Op. cit

[15] See. GISAID: Global initiative on sharing all influenza data (2021). "Open access to epidemic and pandemic virus data." Recovered from: https://www.gisaid.org/

[16] See. GISAID: Global initiative on sharing all influenza data (2021). Open access to epidemic and pandemic virus data". Op. cit.

[17] See. GISAID: Global initiative on sharing all influenza data (2021). Op. cit.

[18] See. GISAID: Global initiative on sharing all influenza data (2021). Op. cit.

[19] See. GISAID: Global initiative on sharing all influenza data (2021). Op. cit.

[20] See. GISAID: Global initiative on sharing all influenza data (2021). Op. cit.

Graph 2
Genomic variants for (SARS-CoV-2) COVID
identified in Mexico between March 2020 and September 2021

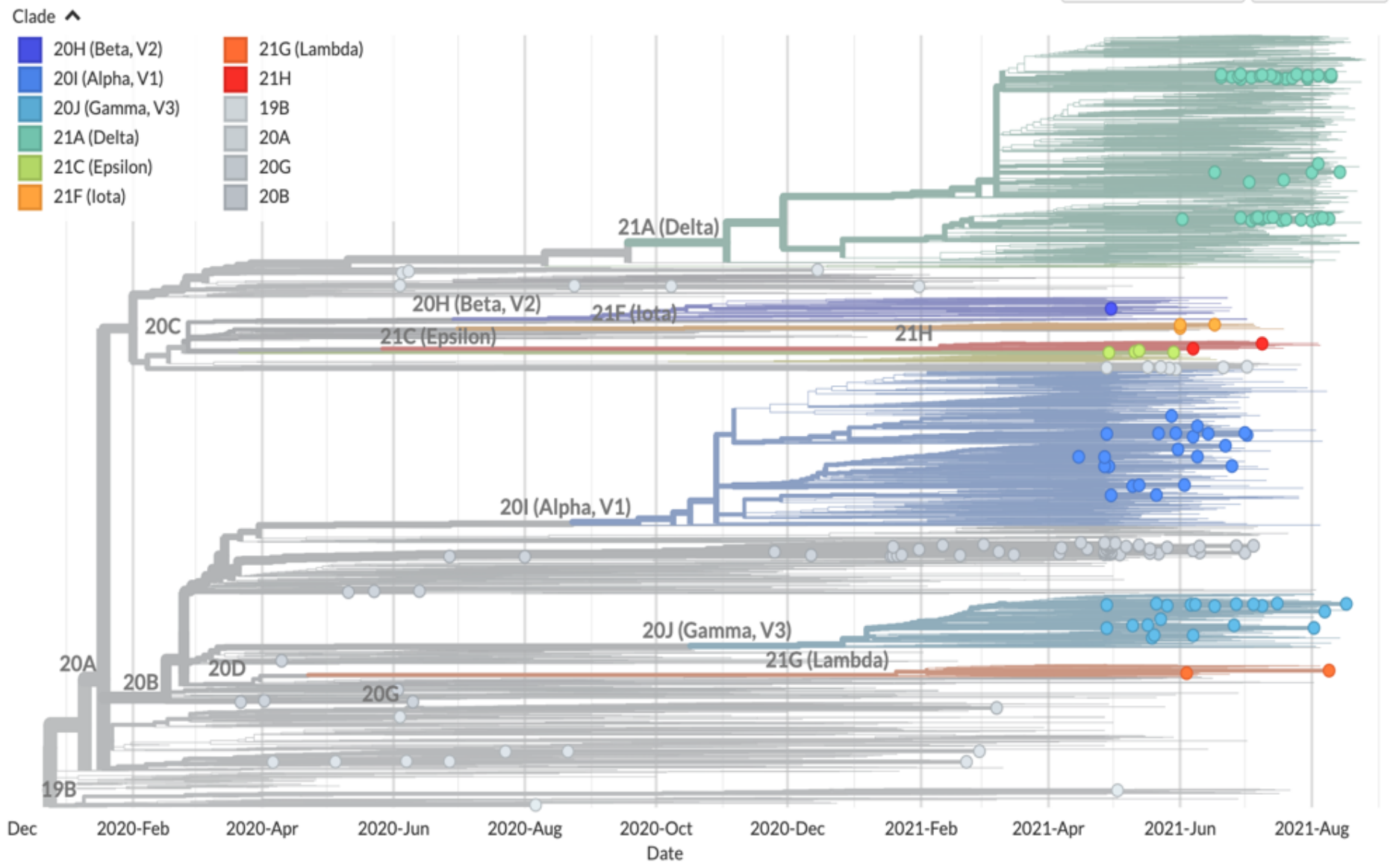

Source: NextTrain. (2021). "Genomic epidemiology of novel coronavirus. Global subsampling". September 3, 2021. Retrieved from: https://nextstrain.org/ncov/gisaid/global

Graph 2 shows the distribution of 435 genomes registered in the GISAID, which were detected in Mexico between March 2020 and September 2021. It should be noted that 85 records of the genomic variant were found *[VOC], [**Delta** (WHO label)], [**G/478K.V1** (GISAID clade) ],*[**B.1.617.2+AY.x** (Pango linaje)] in the country. The variant ***Delta*** *(WHO label)* it is one of the variants that has shown a higher level of contagion.

## Data value

The (WHO, 2020) describes that the pattern of (SARS-CoV-2) COVID-19 presents four typologies depending on the severity of the intensity of contagion, these being: i) not infected or ii) infected; in this finally, four categories are identified in each case: a) mild infection, b) moderate infection, c) severe infection and d) critical infection.

Depending on the category observed in patients who are confirmed infected, as in the case of a) or b), it can assume the outpatient nature, so the strategy is isolation or “quarantine” at home, where the expected result is he recovers. Regarding patients with confirmed infection, in categories c) and d) they assume the character of hospitalized patients, with the probability of requiring care in intensive care units and requiring intubation, and where it is expected to save as many patients as possible , in the face of scarce hospital resources and a growing demand from patients who require specialized medical care services.

In (Alvarez-Luna *et al.,* 2021) and (Dogan, O., Tiwari, S., Jabbar, MA *et al.,* 2021) agree that the study of (SARS-CoV-2) COVID 19 has been addressed by few studies that have used data mining analysis techniques, highlighting the works presented by (Albahri AS, Hamid RA *et al*., 2020), (Ding Z, Qin Z, Qin Z, 2017), (Kumar S, 2020 ), (Medel-Ramírez C, Medel-Lopez H, 2020), (Melin, et al., 2020) and (Muik, *et al.,* 2021) and (Wahbeh A, Nasralah T, Al-Ramahi M, El- Gayar O, 2020).

(Medel-Ramírez C, Medel-Lopez H, 2020) present a study through the application of data mining techniques and the design of an algorithm for the identification of indigenous language speaking patients, who have presented a positive result for (SARS -CoV-2) COVID 19 in Mexico. This study analyzes the information at the municipal, state and national level, presenting the categories of analysis related to: age, sex, present comorbidities, as well as the medical strategy used in the treatment and accounts for the deaths registered according to their degree of infection. presented by the patients.

**Fuente de datos**

This study draws on two sources of information: a) At the national level, with information from the Epidemiological Surveillance System for Viral Respiratory Diseases (SVEERV), in charge of the Ministry of Health of Mexico; and b) At an international level, with information from the Information System of the Global Initiative on Sharing All Influenza Data (GISAID) based in Germany.

The Ministry of Health with the participation of the National Council of Science and Technology (CONACYT), the Center for Research in Geospatial Information Sciences (CENTROGEO), the National Geointelligence Laboratory (GEOINT), the Data Laboratory of the National Geointelligence Laboratory (DataLab), present a daily report that accounts for the presence of cases of (SARS-CoV-2) COVID-19, being this official means of communication and information on the epidemic in Mexico.[21] The information in the SVEERV is disseminated as open data and its information is presented at the municipal, state and national levels. This source of information provides the number of cases registered for (SARS-CoV-2) COVID-19 by the Ministry of Health responsible for the Epidemiological Surveillance System for Viral Respiratory Diseases (SVEERV) and comes from the website https://data.gob.mx/busca/dataset/informacion-referente-a-canes-covid-19-en-mexico

The second source of information is the Global Initiative on Sharing All Influenza Data (GISAID) Information System, which reports on the genomic monitoring and surveillance of (SARS-CoV-2) COVID-19, by registering and identifying variants and mutations. reported by public health organizations, diagnostic laboratories, and international public and private health institutions.

## Method

The information processing is carried out through the design of an algorithm[22] which is built with the support of Orange Data Mining application software for data mining and visual programming [23] version 3.29.3., which is a mining suite for data analysis that uses visual programming and Python scripts. In this way, the following algorithms have been designed for the study of (SARS-CoV-2) COVID 19 in Mexico: a) algorithm for estimating the fatality rate and positivity index in registered cases of (SARS-CoV2) COVID 19 that indicated speaking an indigenous language in Mexico, b) an algorithm for estimating the severity rate of the (SARS-CoV2) COVID 19 infection in the indigenous language-speaking population in Mexico; and c) algorithm for genomic analysis and monitoring of VOC ij Delta = [Delta (WHO label)], [G / 478K.V1 (GISAID clade)], [B.1.617.2 + AY.x (Pango lineage)] of the (SARS-CoV-2) COVID-19 in Mexico.

[21] The information of the registered cases of (SARS-CoV-2) COVID-19 in Mexico, is concentrated daily in the Epidemiological Surveillance System of Viral Respiratory Diseases since April 19, 2020, this being the communication channel and official information on the epidemic in Mexico. The data are presented with a disaggregation at the municipal, state and national levels, taking into account the categories of age, sex, present comorbidities and associated with the presence of (SARS-CoV-2) COVID-19.

[22] The design of the data mining algorithm presented below is made up of the definitions, heiristic conditions and calculations by which the model for the analysis of the information is built to know the impact of the (SARS-CoV-2) COVID-19) in Mexico. The structure is presented in pseudo code, which uses visual programming and Python scripts.

Next, the sections that give an account of the methodological structure for the construction of the algorithms using data mining techniques for the analysis of the impact and genomic monitoring of (SARS-CoV-2) COVID-19 in Mexico are presented.

## 11. Analysis of the impact of (SARS-CoV-2) COVID-19 on the indigenous language-speaking population in Mexico

With the data source that comes from the information registered by the Ministry of Health of Mexico, and that is presented as open data (open-data), in comma separated values (CVS) format. The data processing corresponds to the records of the COVID-19 epidemic (SARS-CoV-2) with a cutoff to September 3, 2021. In the construction of the algorithm for the study of (SARS-CoV-2) COVID-19 In Mexico, an information analysis model is based, which is structured based on the following definitions.

[23] The Orange Data Mininig application software for data mining and visual programming was developed by the Bioinformatics Laboratory of the Faculty of Informatics and Information Sciences, University of Ljubljana, Slovenia. It is free software and its main characteristics lie in its functionality for front-end visual programming aimed at data exploration and results visualization, as well as a specialized Python library. See. Orange Data Mining software version 3.29.3. Recovered from: https://orangedatamining.com/

**Definition a.1.-** Total patients to be considered in the model (SARS-CoV-2) COVID-19.- It is the number of total patients according to the confirmatory laboratory result for (SARS-CoV-2) COVID-19).

Be:

$TP^{SARS\text{-}CoV\text{-}2\ i\ j}$ = Total patients according to (SARS-CoV-2) COVID-19 according to confirmatory result.

Of which:

$P^{+\ SARS\text{-}CoV\text{-}2\ i\ j}$ = Total patients with a positive result (SARS-CoV-2) COVID-19 in the state and / or municipality.

$P^{-\ SARS\text{-}CoV\text{-}2\ ij}$ = Total patients with a negative result (SARS-CoV-2) COVID-19 in the state and / or municipality.

$P^{x\ SARS\text{-}CoV\text{-}2\ i\ j}$ = Total suspected patients[24] (SARS-CoV-2) COVID-19 in the state and / or municipality.

Then:

$$TP^{SARS\text{-}CoV\text{-}2\ I\ j} = (P^{+\ SARS\text{-}CoV\text{-}2\ I\ j)} + (P^{-\ SARS\text{-}CoV\text{-}2\ I\ j)} + (P^{x\ SARS\text{-}CoV\text{-}2\ I\ j})$$

donde: I = state, j = municipality.

**Definition a.2.-** Total number of patients with positive confirmation for (SARS-CoV-2) COVID-19 according to medical treatment strategy (MTE).- Corresponds to the total number of patients with positive confirmation for SARS-CoV-2 COVID- 19, according to the medical treatment strategy determined for their care, based on the type of severity of infection and present comorbidities.

[24] According to the World Health Organization, there are 3 categories (identified as Type 1, Type 2 and Type 3) to identify suspected cases of (SARS-CoV-2) COVID-19. A patient assumes the category of suspect for (SARS-CoV-2) COVID-19 if the diagnostic characteristics indicated by the World Health Organization present the following considerations. Let: Px SARS-CoV-2 ij = Patient with initial classification as a suspected case of (SARS-CoV-2) COVID-19, where: Px SARS-CoV-2 ij = Px (SARS-CoV-2) COVID-19 Type 1 + Px (SARS-CoV-2) COVID-19 Type 2 + Px (SARSCoV-2) COVID-19 Type 3. Identifying the following scenarios:

1. Px (SARS-CoV-2) COVID-19 Type 1.- Confirmed by association applies when the case reported being a positive contact for COVID-19 (and this is registered in the SISVER) and the case is not I took a sample or the sample was invalid.

2. Px (SARS-CoV-2) COVID-19 Type 2.- Confirmed by ruling only applies to deaths under the following conditions: The case was not sampled or a sample was taken, but the sample was invalid.

3. Px (SARS-CoV-2) COVID-19 Type 3.- Confirmed applies when: The case has a laboratory sample or antigenic test and was positive for SARS-CoV-2, regardless of whether the case has a clinical epidemiological association.

The medical care strategy required for patients with a positive result for (SARS-CoV2) COVID-19, according to their identified severity of infection, presents the following medical care scenarios:

Be:

$ETM^{P+\ SARS\text{-}CoV\text{-}2}{}_{i\,j}$ = Medical treatment strategy for patients with a positive result for (SARS-CoV-2) COVID-19, based on the medical treatment strategy (ETM P + SARS-CoV-2 ij), according to their degree of infection and present comorbidities , proposes two action scenarios: 1) Health care strategy for patients with a positive result for (SARS-CoV2) COVID-19 with outpatient medical care, and 2) Health care strategy for patients with a positive result for (SARS-CoV2) COVID-19 with hospital medical care.

Be:

1. $ETM^{Ambulatory\ [\ P+\ SARS\text{-}CoV\text{-}2\ _{i\,j}\ ]}$ = Total number of patients with a positive result for (SARS-CoV2) COVID-19 with an outpatient medical care strategy, by state and / or municipality.
2. $ETM^{Hospitable\ [\ P+\ SARS\text{-}CoV\text{-}2\ i\ j\ ]}$ = Total patients with a positive result for (SARS-CoV2) COVID-19 with hospital medical care [25] according to state and / or municipality.

Then:

$$ETM^{P+\ SARS\text{-}CoV\text{-}2}{}_{I\,j} = ETM^{Ambulatory\ [\ P+\ SARS\text{-}CoV\text{-}2\ _{I\,j}\ ]} + ETM^{Hospitable\ [\ P+\ SARS\text{-}CoV\text{-}2\ I\ j\ ]}$$

**Definition a.3.-** Total number of patients with a positive result for (SARS-CoV-2) COVID-19 with hospital medical care with access to the intensive care area (ICU) .- It is the total number of patients with a positive result for SARS-CoV-2 COVID- 19 with medical treatment strategy ($ETM^{Hospitalary\ [\ P+\ SARS\text{-}CoV\text{-}2\ i\ j\ ]}$) that require access to the intensive care unit (ICU), depending on the degree of severity of the infection.

Be:

[25] The care of P + SARS-CoV-2 ICU i j, depending on the severity of the infection, can access two care scenarios:

I. Patients with a positive result for (SARS-CoV-2) COVID-19 with hospital medical care, with access to the intensive care area (ICU) and with access to an intubation procedure for respiratory failure.

II. Patients with a positive result for (SARS-CoV-2) COVID-19 with hospital medical care, with access to the intensive care area (ICU), who do not require intubation.

[ ( ETM $^{\text{Hospitalary ( P+ SARS-CoV-2 i j ) UCI}}$ ] $_{i\,j}$ = Total number of patients with a positive result for (SARS-CoV-2) COVID-19 with a hospital medical care strategy and with access to the intensive care unit ($^{\text{ICU}}$) according to state and / or municipality.

**Definition a.4.-** Total patients with a positive result for (SARS-CoV-2) COVID-19 with hospital medical care with access to the intensive care unit ($^{\text{ICU}}$) and with an intubation procedure for respiratory failure in the intensive care area.

Be:

[ ( ETM $^{\text{Hospitaly ( P+ SARS-CoV-2 ) ICU (Intubation)}}$ ] $_{i\,j}$ = Total number of patients with a positive result for (SARS-CoV-2) COVID-19 with a hospital medical care strategy with access to an intensive care unit and with an intubation procedure for respiratory failure, according to state and / or municipality.

**Definition a.5.-** Total number of deaths of patients with a positive result for (SARS-CoV-2) COVID-19.- Corresponds to the total number of deaths of patients with a positive confirmation for (SARS-CoV-2) COVID-19 registered in the System database of Epidemiological Surveillance of Viral Respiratory Diseases.

Be:

Deaths $^{[\text{P+SARS-CoV-2}_{ij}]}$ = Total number of deaths of patients with positive confirmation for (SARS-CoV 2) COVID-19 registered in the database of the Epidemiological Surveillance System for Viral Respiratory Diseases, and whose value in the field (DATE_DEF is different from the value “9999- 99-99 ").

**Definition a.6.-** Fatality rate (SARS-CoV-2) COVID-19.- It is the proportion of the number of people killed by (SARSCoV-2) COVID-19 among the total number of patients with a positive result for (SARS-CoV-2) COVID- 19, in a period and in a determined area.

Be:

TL $^{\text{SARS-CoV-2}}{}_{i\,j}$ = Fatality rate of (SARS-CoV-2) COVID-19 by state and / or municipality.

Where:

1. Deaths $^{[P+\ SARS\text{-}CoV\text{-}2}\ _{ij}{}^{]}$ = Total number of deaths of patients with positive confirmation for (SARS-CoV 2) COVID-19 registered in the database of the Epidemiological Surveillance System for Viral Respiratory Diseases according to state and / or municipality.
2. $P^{+\ SARS\text{-}CoV\text{-}2\ i\ j}$ = Total patients with a positive result (SARS-CoV-2) COVID-19 in the state and / or municipality.

Being:

$TL^{\ SARS\text{-}CoV\text{-}2}{}_{i\ j}$ = (Total number of deaths of patients with positive confirmation for (SARS-CoV 2) COVID-19 registered in the database of the Epidemiological Surveillance System for Viral Respiratory Diseases according to state and / or municipality divided by the total number of patients with a positive result (SARS-CoV-2) COVID-19 in the state and / or municipality) per 100.

Then:

$$TL^{\ SARS\text{-}CoV\text{-}2}{}_{I\ j.} = [\ (Deaths^{\ [P+\ SARS\text{-}CoV\text{-}2}\ _{ij}{}^{]}) \ /\ P+\ SARS\text{-}CoV\text{-}2_{\ i\ j}\ ] \times 100$$

**Definition a.7.-** Index of positivity (SARS-CoV-2) COVID-19.- It is the proportion of the number of patients with a positive result (SARS-CoV-2) COVID-19 and the total of samples (positive plus negative) for an epidemiological week of onset of symptoms, in a federal entity determines residence.

Be:

$IP^{\ SARS\text{-}CoV\text{-}2}{}_{I\ j.} =$ COVID-19 (SARS-CoV-2) positivity index

Where:

$$IP^{\ SARS\text{-}CoV\text{-}2}{}_{I\ j.} = [\ (P^{+\ SARS\text{-}CoV\text{-}2\ i\ j}) \ /\ (P^{+\ SARS\text{-}CoV\text{-}2\ i\ j}) + (P^{-\ SARS\text{-}CoV\text{-}2\ ij})] \times 100$$

Being:

$P^{+\ SARS\text{-}CoV\text{-}2\ ij}$ = Total patients with a positive result (SARS-CoV-2) COVID-19 in the state and / or municipality.

$P^{-\ SARS\text{-}CoV\text{-}2\ ij}$ = Total patients with a negative result (SARS-CoV-2) COVID-19 in the state and / or municipality.

The data treatment corresponds to the records on the COVID-19 epidemic (SARS-CoV-2) with information at the cutoff corresponding to September 3, 2021. The information is processed through the application software for mining data Orange version 3.29.3, in with the development of the algorithm for the analysis of the information. (See Figure 1, below). According to information from the Government of Mexico through the information registered by the Ministry of Health, and whose records are available in its open data modality.

**Definition a.8.-** Rate of severity of infection in patients with a positive result for (SARS-CoV-2) COVID-19.- Corresponds to the identification of the degree of infection in patients with a positive result for (SARS-CoV-2) COVID-19, according to symptoms, comorbidities, and hospital medical strategy used for their care.

Be:

$TGI^{SARS\text{-}CoV\text{-}2}_{I\,j.} =$ Infection severity rate in patient with confirmation positive for (SARS-CoV-2) COVID-19

Where:

$$TGI^{SARS\text{-}CoV\text{-}2}_{I\,j.} = [\ TGI^{1}_{I\,j} + TGI^{2}_{I\,j} + TG^{3}_{I\,j}\ ]$$

Being:

$TGI^{1}_{I\,j}$ = Type of mild infection in patients with positive confirmation for (SARS-CoV-2) COVID-19 in the state and / or municipality.

$TGI^{2}_{I\,j}$ = Moderate infection type in patients with positive confirmation for (SARS-CoV-2) COVID-19 in the state and / or municipality.

$TGI^{3}_{I\,j}$ = Severe infection type in patients with positive confirmation for (SARS-CoV-2) COVID-19 in the state and / or municipality.

Which one:

$TGI^{1}_{I\,j}$ = Type of mild infection in patients with positive confirmation for (SARS-CoV-2) COVID-19 in the state and / or municipality.- Represents the proportion of the total number of patients with a positive result for (SARS-CoV-2) COVID- 19 and who have received an outpatient health care strategy among the total number of with a positive result for (SARS-CoV-2) COVID-19, at the national, state and / or municipal level.

Be:

$$TGI^{1}_{I\,j} = [\ ETM^{Ambulatory\ [P+\ SARS\text{-}CoV\text{-}2}{}_{i\,j} / P^{+\ SARS\text{-}CoV\text{-}2\ i\,j}\ ] \times 100$$

$TGI^{2}_{I\,j}$ = Moderate infection type in patients with positive confirmation for (SARS-CoV-2) COVID-19 in the state and / or municipality.- Represents the proportion of the total number of patients with positive result for (SARS-CoV-2) COVID- 19 and who have received a hospital medical care strategy minus the total number of patients with a positive result for (SARS-CoV-2) COVID-19 with a hospital medical care strategy and with access to the intensive care unit (ICU) according to state and / or municipality, divided by the total number of with a positive result for (SARS-CoV-2) COVID-19, at the national, state and / or municipal level.

Be:

$$TGI^{2}_{I\,j} = [\ (ETM^{\text{Hospitalary [ P+ SARS-CoV-2}}{}_{i\,j}) - (ETM^{\text{Hospitalary ( P+ SARS-CoV-2 ) UCI (Intubation)}}{}_{i\,j}) \ / \ P^{+\ \text{SARS-CoV-2}}{}_{i\,j}\ ] \times 100$$

$TGI^{3}_{I\,j}$ = Type of severe infection in patients with positive confirmation for (SARS-CoV-2) COVID-19 in the state and / or municipality.- Represents the proportion of the total number of patients with a positive result for (SARS-CoV-2) COVID-19 and who have received a hospital medical care strategy and with access to the unit of intensive care (ICU) by state and / or municipality, divided by the total number of with a positive result for (SARS-CoV-2) COVID-19, at the national, state and / or municipal level.

Be:

$$TGI^{3}_{I\,j} = [\ ETM^{\text{Hospitalary ( P+ SARS-CoV-2 ) UCI (Intubation)}}{}_{i\,j}) \ / \ P^{+\ \text{SARS-CoV-2 i j}}\ ] \times 100$$

Figure 1
Algorithm for estimating the fatality rate and positivity index
in recorded cases of (SARS-CoV2) COVID 19
who indicated that they spoke an indigenous language in Mexico

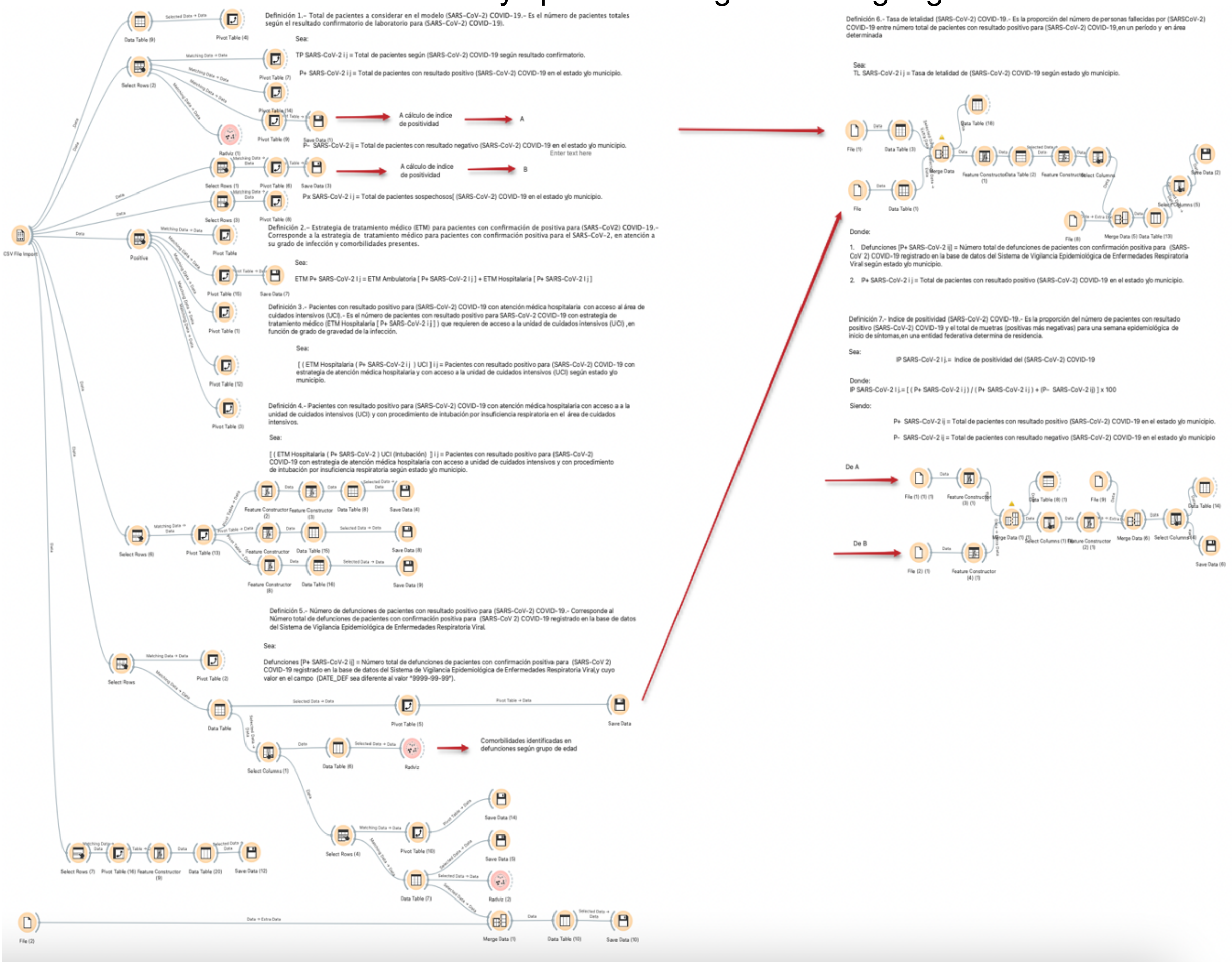

Source: self made. Using visual data mining scripts in Orange Data Mining version 3.29.3

Figure 2
Algorithm for estimating the severity rate of COVID 19 (SARS-CoV2) infection in the indigenous language-speaking population in Mexico

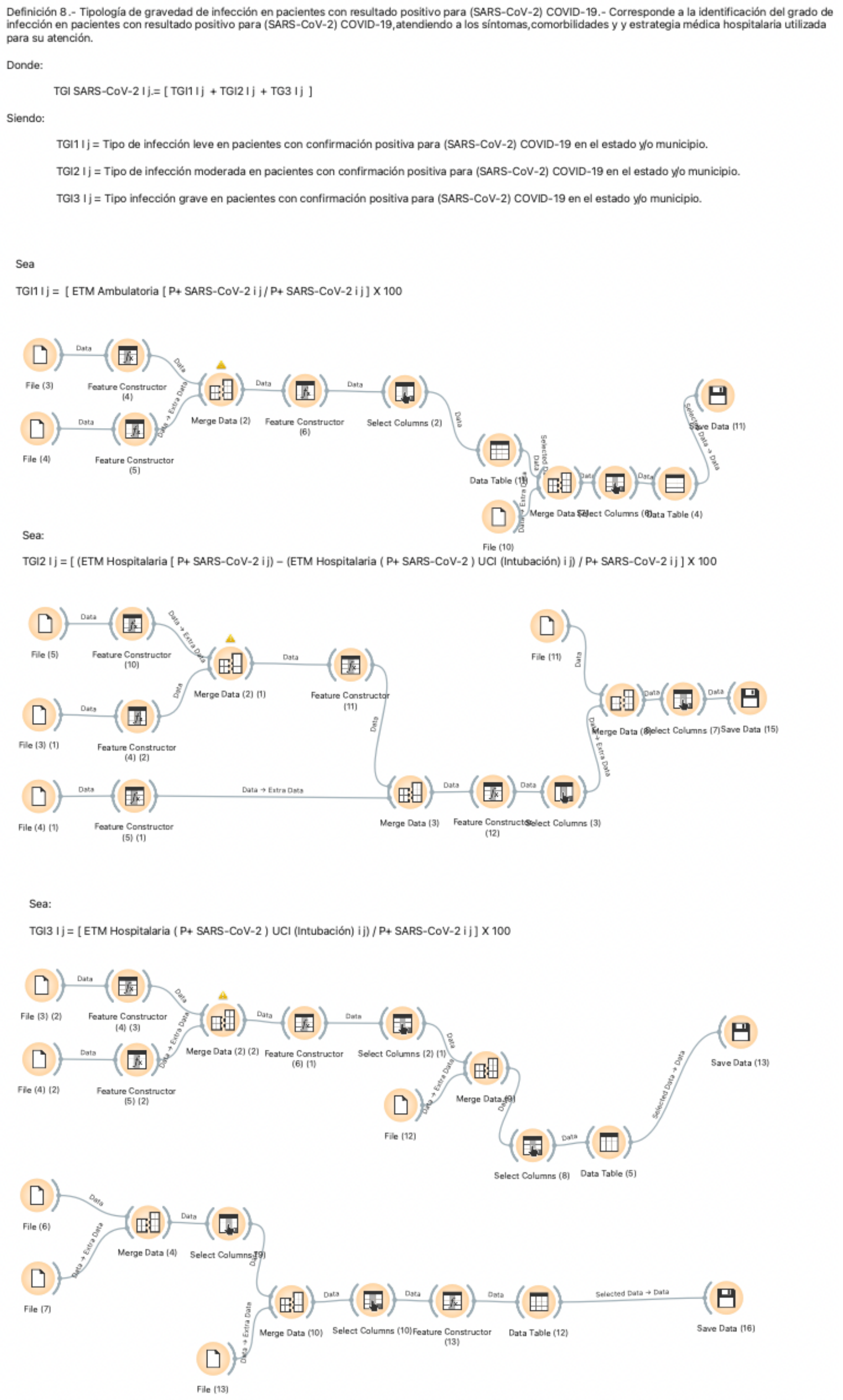

Source: self made. Using visual data mining scripts in Orange Data Mining version 3.29.3

## 12. Genomic monitoring and surveillance of (SARS-CoV-2) COVID-19 in the population in Mexico

**Definition b.1.-** Genomic variant of (SARS-CoV-2) identified COVID-19.- Corresponds to the identification of complete sequences of the viral genome associated with SARS-CoV-2) COVID-19. [26]

Be:

$$VGID_{ij}^{\text{(SARS-CoV-2) COVID-19}} = VOC_{ij}^{Alpha} + VOC_{ij}^{Beta} + VOC_{ij}^{Gamma} + VOC_{ij}^{Delta} + VOI_{ij}^{Eta} + VOI_{ij}^{Iota} + VOI_{ij}^{Kappa} + VOI_{ij}^{Lambda} + VOI_{ij}^{Mu}$$

Where:

$VOC_{ij}^{Alpha}$ = [GRY (GISAID clade)], [ B.1.1.7+Q.x (Pango lineage)],

$VOC_{ij}^{Beta}$ = [Beta (WHO label)], [GH/501Y.V2 (GISAID clade)], [ B.1.351 + B.1.351.2 + B.1.351.3 (Pango lineage) ]

$VOC_{ij}^{Gamma}$ = *[Gamma (WHO label) ], [GR/501Y.V3 (GISAID clade) ],* [ P.1+P.1.x (Pango lineage) ]

$VOC_{ij}^{Delta}$ = *[Delta (WHO label) ], [G/478K.V1 (GISAID clade) ],*[B.1.617.2+AY.x (Pango lineage)]

$VOI_{ij}^{Eta}$ = *[Eta (WHO label) ], [G/484K.V3 (GISAID clade)], [ B.1.525* (Pango lineage) ]

$VOI_{ij}^{Iota}$ = *[Iota (WHO label) ], [ HG/253G.V1 (GISAID clade) ], [B.1.526* (Pango lineage) ]

$VOI_{ij}^{Kappa}$= *[Kappa (WHO label) ], [G/452R.V3 (GISAID clade) ], [B.1.617.1* (Pango lineage]

$VOI_{ij}^{Lambda}$ = *[Lambda (WHO label) ], [GR/452Q.V1 (GISAID clade) ], [C.37* (Pango lineage) ]

$VOI_{ij}^{Mu}$ = *[Mu (WHO label) ], [GH(GISAID clade) ], [B.1.121+b.1.21.1* (Pango lineage) ]

Being :

[$i$ = State ] , [ $j$ = Municipality ]

Where:

VOC = Variant of concern [27] VOI = Variant of interest [28]

---

[26] See. World Health Organization (2021). Genomic sequencing of SARS-CoV-2. A guide to implementation for maximum impact on public health. Recovered from: https://www.who.int/publications/i/item/9789240018440

[27] In World Health Organization (2021). Op. Cit. it refers to “the variant of concern or VOC which may present the following scenarios: a) it is associated with an increase in transmissibility or a worsening of the epidemiological situation in the region; b) it is associated with an increase in virulence or change in clinical presentation; or c) is associated with a decrease in the effectiveness of control measures, diagnostic tests, vaccines or treatments. " See. Argentine Ministry of Health (2021). "Integration of SARS-CoV-2 genomic surveillance to covid-19 surveillance through the National Health Surveillance System. Version 1. 04/2021." Retrieved from: https://bancos.salud.gob.ar/sites/default/files/2021-04/SNVS_integracion-de-la-vigilancia-genomica_de_SARS-CoV-2.pdf

[28] The Ministry of Health of Argentina defines that VOI corresponds to “… a SARS-CoV-2 isolate is a VOI if it phenotypically behaves differently from a reference or its genome presents mutations that lead to amino acid changes associated with established phenotypic changes”, and “If multiple COVID cases, clusters, or has been detected in multiple countries have been

Figure 3
Algorithm for genomic analysis and monitoring of VOC ij Delta = [Delta (WHO label)], [G / 478K.V1 (GISAID clade)], [B.1.617.2 + AY.x (Pango lineage)] del (SARS -CoV-2) COVID-19 in Mexico

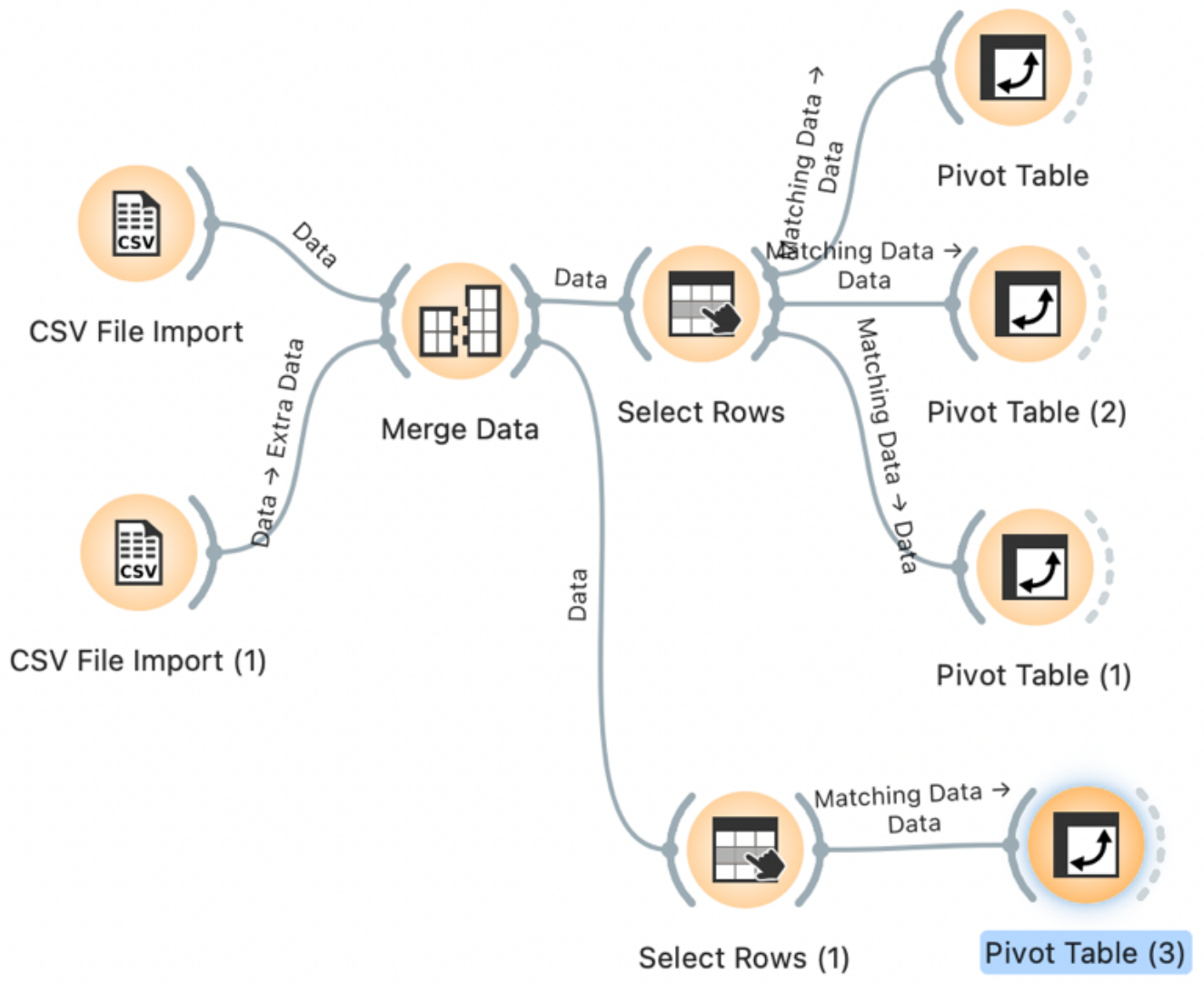

Source: self made. Using visual data mining scripts in Orange Data Mining version 3.29.3

In the genomic monitoring of the variants of (SARS-CoV-2) COVID-19 in Mexico, it is highlighted that as of September 3, 2021, the variant of VOC $_{i\,j}$ $^{Delta}$ = *[Delta (WHO label)], [G/478K.V1 (GISAID clade)],*[B.1.617.2+AY.x (Pango lineage)] it is the one with the largest presence in the country, according to information registered in the Information System of the Global Initiative on Sharing All Influenza Data (GISAID). With the algorithm of Figure 3, the analysis of the presence of the study variant is carried out, and particularly, in the states that showed a high fatality rate, a high index of positivity; Being the State of Puebla, State of Hidalgo, the State of Veracruz de Ignacio de la Llave and the State of Oaxaca, the entities selected to analyze the presence of the delta variant of (SARS-CoV-2) COVID-19.

identified in community circulation." See. See. Argentine Ministry of Health (2021). Integration of the genomic surveillance of SARS-CoV-2 to the surveillance of covid-19 Through the National Health Surveillance System. Op. Cit.

Graph 3 shows that the VOC Delta and VOC Gamma variants have an important presence in the national territory. It should be noted that the VOC Delta variant is one of the most variants that presents a higher level of contagion, and therefore, it is more careful.

Graph 3
Summary of the main variants and mutations of COVID-19 (SARS-CoV-2) according to the proportion identified in Mexico
in the period April 2020 to August 2021

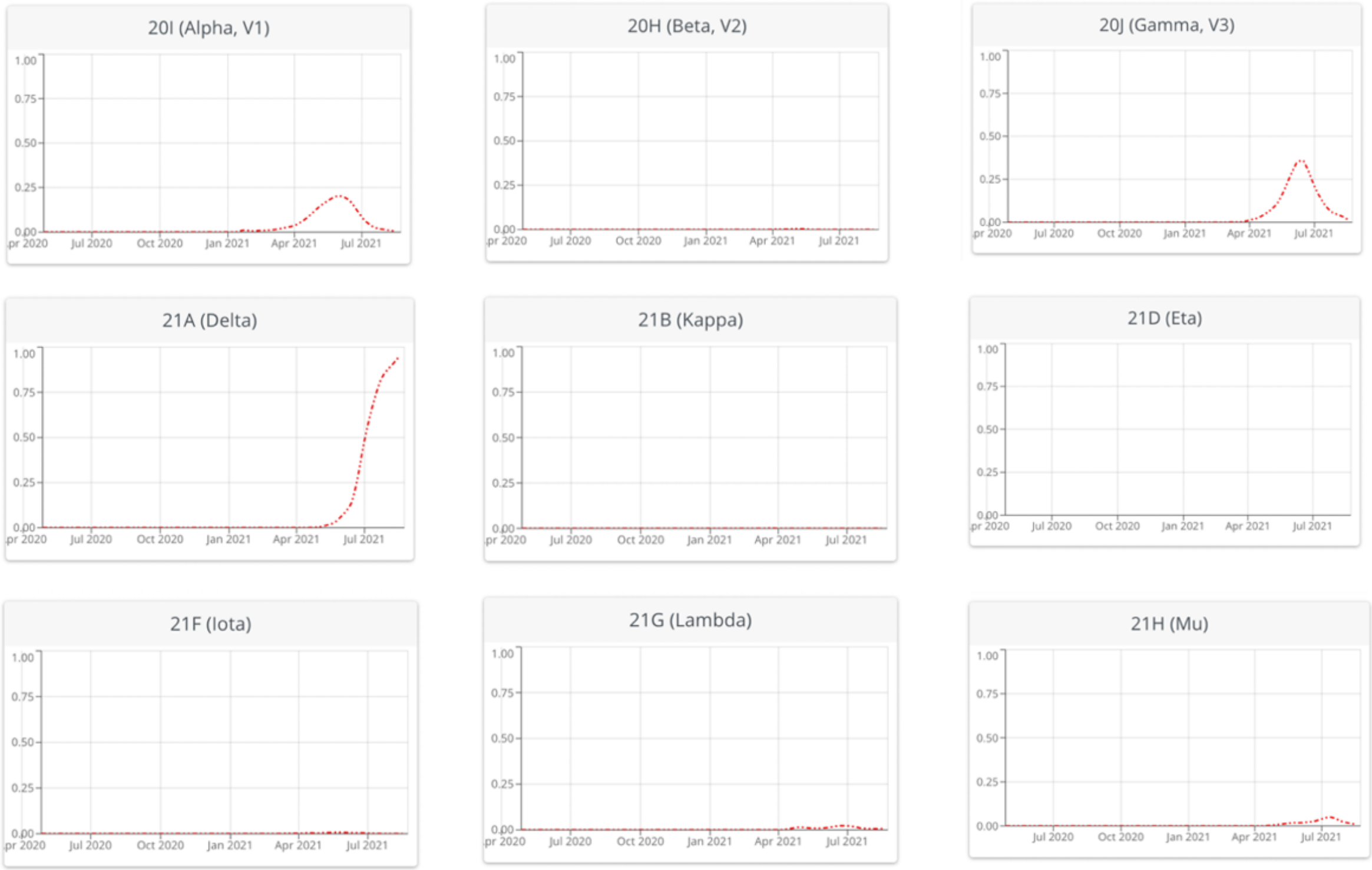

Source: Taken from Institute of Social and Preventive Medicine University of Bern, Switzerland & SIB Swiss Institute of Bioinformatics, Switzerland (2021). With information as of September 3, 2021 Retrieved from: https://covariants.org/per-variant

As of September 3, 2021, in Mexico, 5,495 study samples have been registered for the genomic study of (SARS-CoV-2) COVID-19, of which 51.21% of the records correspond to the variant $VOC_{ij}^{Delta}$ = *[Delta (WHO label) ], [G/478K.V1 (GISAID clade) ]*,[B.1.617.2+AY.x (Pango lineage)] therefore, it is the genomic mutation with the greatest presence in the cases of study and genomic follow-up observed in the country. In second place with 26.81% the variant $VOC_{ij}^{Gamma}$ = *[Gamma (WHO label) ], [GR/501Y.V3 (GISAID clade) ]*, [ P.1+P.1.x (Pango lineage) ] and in third place with 17.52% corresponds to the variant $VOC_{ij}^{Alpha}$ = [GRY(GISAID clade)], [ B.1.1.7+Q.x (Pango lineage)].

It is highlighted that 2,814 study samples for the genomic study of (SARS-CoV-2) COVID-19, corresponding to the VOC $_{i\,j}^{Delta}$ = *[Delta* $_{(WHO\ label)}$ *], [G/478K.V1* $_{(GISAID\ clade)}$ *],*[B.1.617.2+AY.x $_{(Pango\ lineage)}$], this is classified according to the identification of the Clade in the following way: a) 176 records correspond to the Clade G; b) 2,636 correspond to Clade GK, c) only 1 corresponds to Clade CV, and b) only 1 to Clade 0.

In this sense, within the genomic variant, with the greatest presence in Mexico as of September 3, 2021, is the variant VOC $_{i\,j}^{Delta}$ , of which it shows that 2,636 case studies correspond to Clade GK, representing 93.67% of the cases. It is important to note that a variant VOC $_{i\,j}^{Delta}$ it has shown a higher level of contagion, for which it is necessary and pertinent to carry out a follow-up in the national territory. (See Table 8, at the end of the document).

**Results**

With information from the Ministry of Health of Mexico, from the beginning of the registry the presence of (SARS-CoV-2) COVID-19 in the indigenous language-speaking population in Mexico, until September 3, 2021 at the national level, the following scenario is presented:

1. The total number of registration in the indigenous language speaking population is 58,739 cases, of which: a) 21,294 indigenous language speaking patients registered a positive result for (SARSCoV-2) COVID-19, b) 37,445 patients with negative result for (SARS-CoV-2) COVID-19[29]. (See table 1, at the end of the document).

[29] As of September 3, 2021, the information regarding the number of indigenous language speaking patients at the national level that are registered with a negative result for (SARS-CoV-2) COVID-19, estimated at 37,445, is integrated as follows: a ) 72 cases correspond to the category of records "Invalid by laboratory", that is, it applies when the case does not have a clinical epidemiological association, nor a ruling for (SARS-CoV-2) COVID-19, corresponds to the cases in which it was taken a laboratory sample and it was invalid; b) 506 cases correspond to the category of “Not carried out by the laboratory”, and refers to when the case does not have a clinical epidemiological association, nor a ruling for (SARS-CoV-2) COVID-19, as well as for cases in which it is He took a laboratory sample and it was not processed; and c) 5,348 cases correspond to the category

2. The total number of positive cases for (SARSCoV-2) COVID-19 registered in the indigenous-speaking population in Mexico as of September 3, 2021, correspond to 21,294, cases of which 9,915 are women and 11,379 are men. (See table 2).
3. Of the 21,294 patients registered in the indigenous-speaking population with a positive result for (SARSCoV-2) COVID-19: a) 14,761 have outpatient medical care and b) 6,533 patients received hospital medical care. Of the latter, 2,716 are women and 3,817 are men. (See table 3).
4. 10 states concentrate 72.67% of the indigenous language speaking population that register a positive result for (SARS-CoV-2) COVID-19 that have received hospital care, the State of Yucatan stands out, which concentrates 18.51 % of patients, followed by the State of Oaxaca with 12.60%, the State of Mexico with 9.02%, the State of Puebla with 8.30%, the State of Hidalgo with 7.65%, the State of Quintana Roo with 5.34%, the State of Guerrero with 5.10%, the State of Veracruz with 4.85% and the State of Michoacán de Ocampo with 3.57% (See Table 4).
5. In relation to the total number of hospitalized indigenous language speaking patients who present a positive result for (SARS-CoV-2) COVID-19, who required admission to the intensive care unit, as well as intubation, 777 patients, of which 296 are women and 481 are men. (See Table 5).
6. As of September 3, 2021, 3,321 indigenous language speaking patients with a positive record for (SARS-CoV-2) COVID-19 have died in Mexico, 1,248 women and 2,073 men. (See Table 6).
8. Only 358 deaths corresponded to indigenous language speaking patients with a positive record for (SARS-CoV-2) COVID-19 who required admission to

"Suspicious cases" and applies when the case does not have a clinical-epidemiological association, nor a ruling for (SARS-CoV-2) COVID-19 and no sample was taken, or a laboratory sample was taken. and is pending the result, regardless of another condition; and finally, d) 31,519 cases correspond to the category "Negative for (SARS-CoV-2) COVID-19", which applies when the case: 1.- A laboratory sample was taken and it was: negative for (SARS -CoV-2) COVID-19 or positive for any other respiratory virus (Influenza, RSV, Bocavirus, others) regardless of whether this case has a clinical-epidemiological association or ruling on (SARS-CoV-2) COVID-19, or 2. - An antigenic sample was taken that was negative for (SARS-CoV-2) COVID-19 and the case was not taken from a laboratory sample or confirmed by epidemiological association or by epidemiological clinical opinion. See. Table 1 at the end of the document.

Intensive Care Units (ICU) and Intubation and 2,963 corresponded to (SARS-CoV-2) COVID-19 positive patients who were not in intensive care. (See Table 7).

9. Figure 3 shows three states identified in indigenous language speaking patients with a positive result for (SARS-CoV-2) COVID-19 who have received hospital medical care, in which the main comorbidities are presented, at the national level as of September 3, 2021, according to the following characteristics: a) according to age group, b) comorbidities registered in deaths and c) comorbidities identified in deaths of patients treated in an intensive care unit.

Figure 3
Comorbidities identified in indigenous language speaking patients at the national level according to age group and identified characteristic until September 3, 2021

At the national level, according to age group as of September 3, 2021 a)

In deaths nationwide, according to age group as of September 3, 2021 b )

In deaths in intensive care units nationwide, by age group until September 3, 2021c )

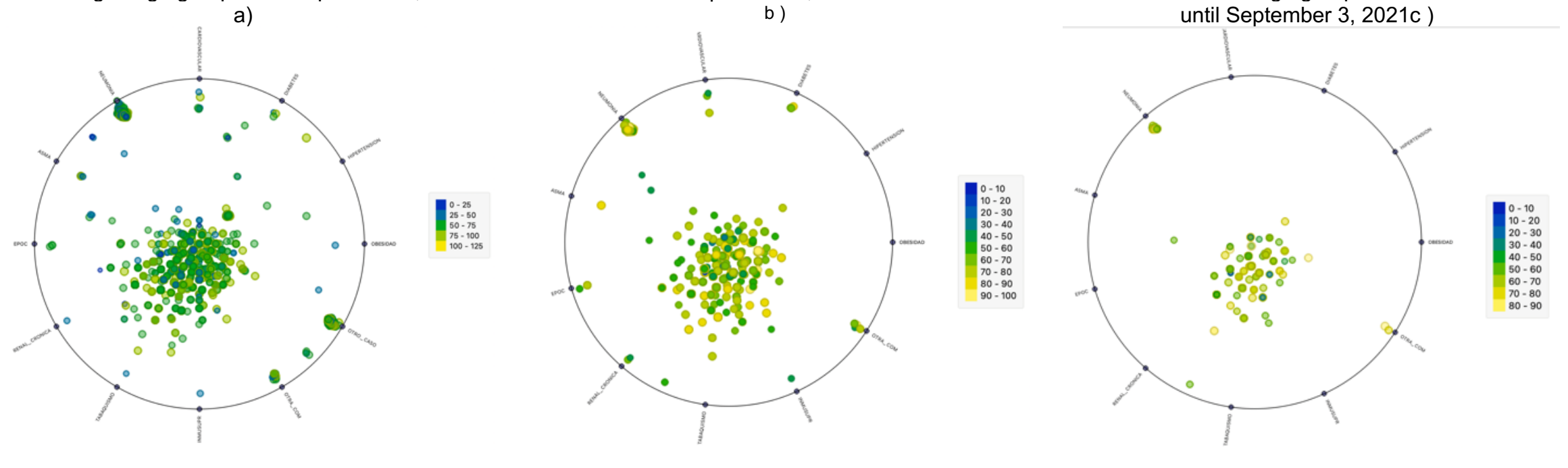

N = 21.294. 9,915 women and 11,379 men.

N = 3,321. 1,248 women and 2,073 men.

N = 358. 122 women and 236 men.

Source: Own elaboration with information as of September 3, 2021.

10. The fatality rate of (SARS-CoV-2) COVID-19 in indigenous language speaking patients, nationwide, as of September 3, 2021 is estimated at 13.5%. It is important to note that six states registered the highest fatality rates, accumulated as of September 3, 2021 are the following: a) Puebla with a rate of 30.5%, b) Campeche with 25.1%, c) Oaxaca with a rate of 20.9%, d) Quintana Roo with a rate of 20.9%, e) Tlaxcala with a rate of 20.0% and f) Veracruz de Ignacio de la Llave with a rate of 19.8%. (See Chart 4, below).

Graph 4
Fatality rate and positivity index of SARS-CoV-2)
COVID-19 in indigenous language-speaking population
according to state in Mexico,
until September 3, 2021

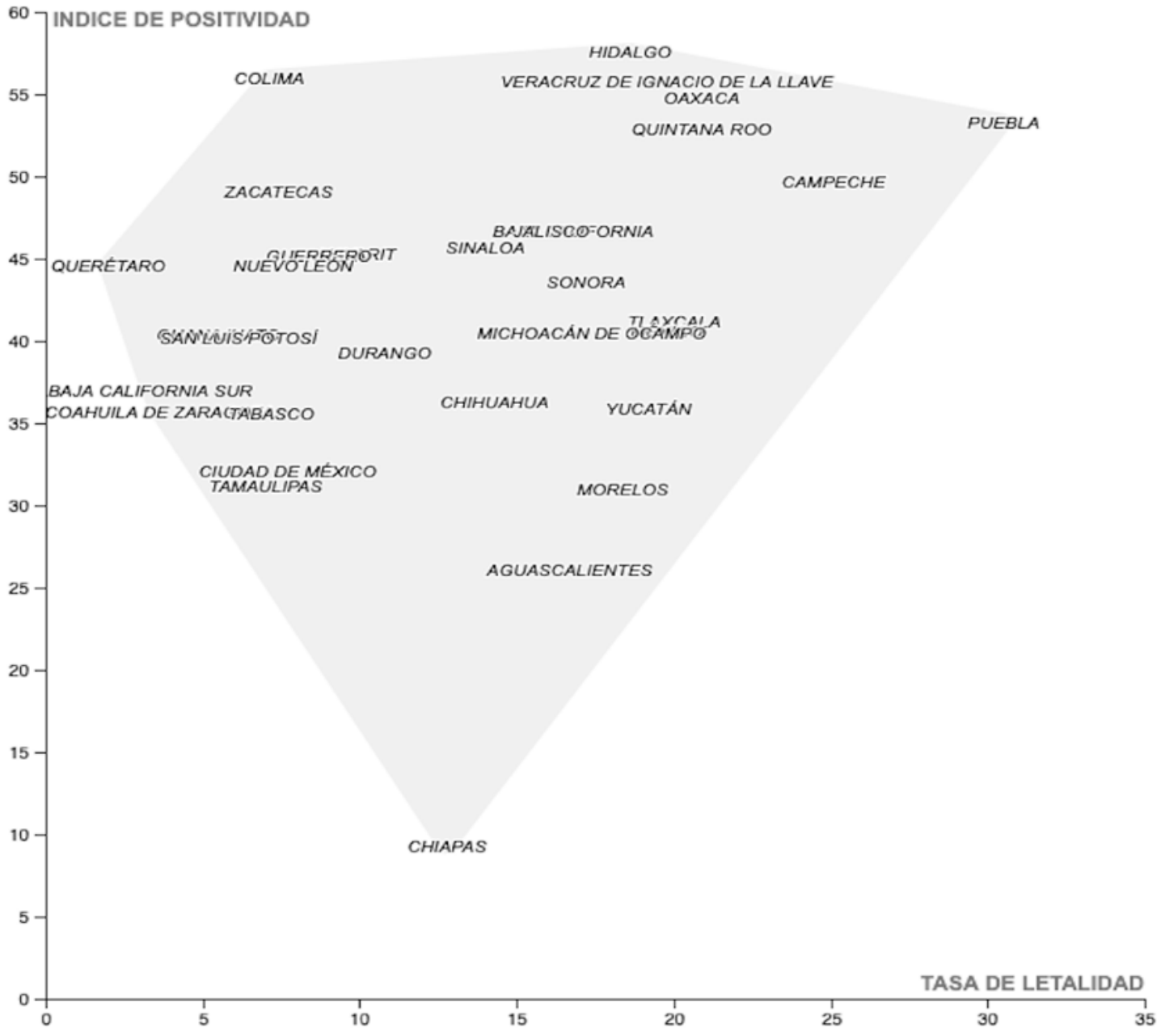

Source: Own elaboration with information as of September 3, 2021.

11. Regarding the positivity index of SARS-CoV-2) COVID-19 in patients who speak indigenous languages, at the national level, as of September 3, 2021, where it is highlighted that the six states that present a higher level of positivity are the following: a) The State of Hidalgo with a positivity index of 57.6%, b) Colima with 56.0%, c) Veracruz de Ignacio de la Llave with 55.8%, d) Oaxaca with 54.8%, e) Puebla with 53.3 % and f) Quintana Roo with 52.9%. (See. Graph 4, presented previously).
12. The severity rate of the SARS-CoV-2) COVID-19 infection present in indigenous language speaking patients in Mexico is made up of three levels: a) the mild infection rate, b) the moderate infection rate and c) the rate of severe infection. See Graph 5, below.

Graph 5
SARS-CoV-2) COVID-19 infection severity rate in indigenous language-speaking population according to state, as of September 3, 2021

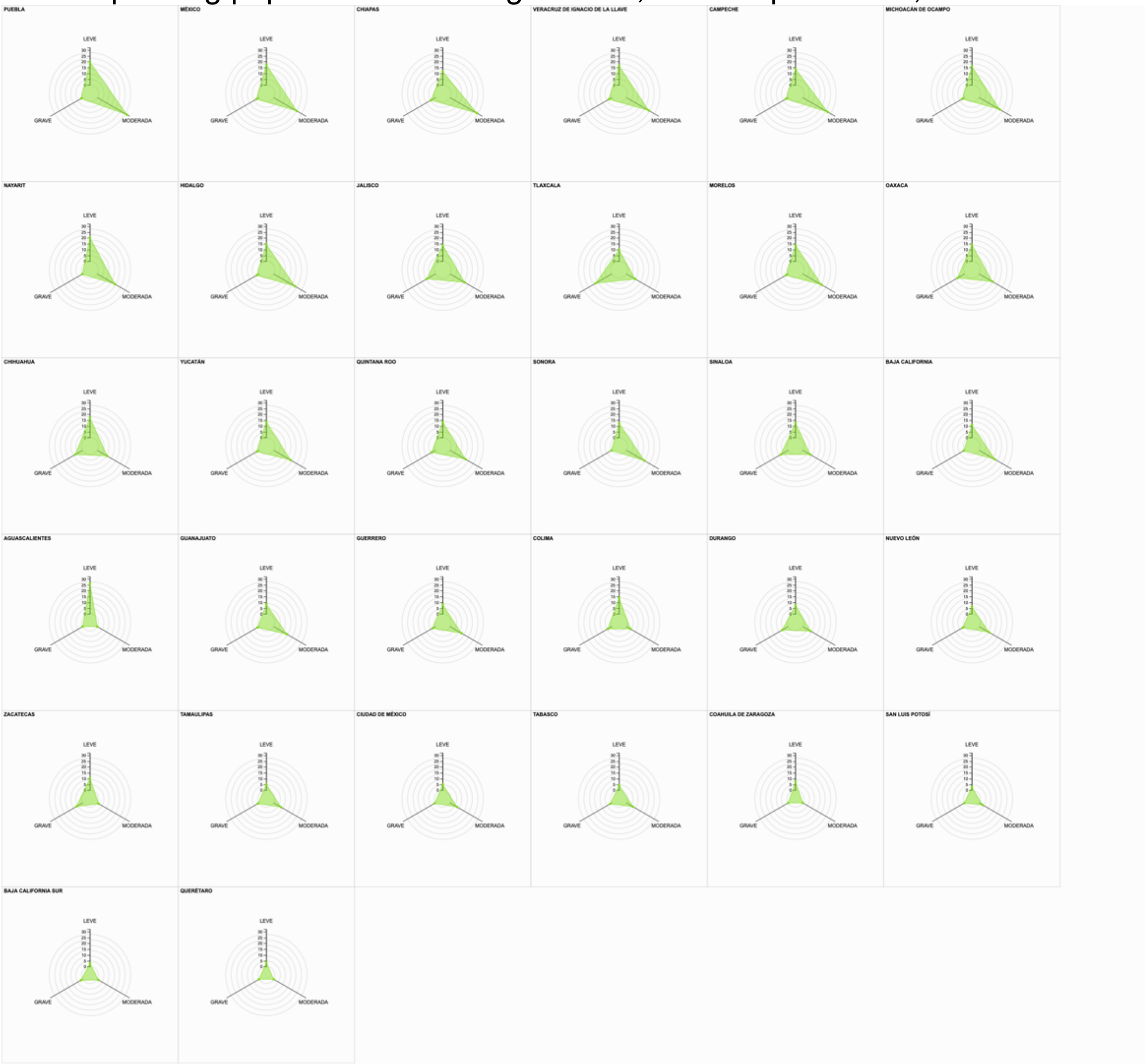

Source: Own elaboration with information as of September 3, 2021.

13. From the information presented in Graph 4 above, it stands out that the five states that registered high rates of severity in infection in patients who speak an indigenous language by (SARS-CoV-2) COVID-19, at the national level and with Accumulated figure as of September 3, 2021, are: a) The State of Puebla with 52.2%, b) State of Mexico with 46.4%, c) State of Chiapas with 46.2%, d) State of Veracruz de Ignacio de la Llave with 45.2% and e) State of Campeche with 43.9%.

14. With information as of September 3, 2021 from the Information System of the Global Initiative on Sharing All Influenza Data (GISAID), regarding the genomic tracking of COVID-19 variants (SARS-CoV-2) present in Mexico, highlights that of 5,495 case studies registered on this platform, 51.21% correspond to the Delta variant, 26.81% correspond to the Gamma variant and 17.52% were identified as the Alpha variant. The presence of new variants of (SARS-CoV-2) COVID-19, such as the Mu variant, was only identified in 2.69% of the total study records in Mexico, at the date established as the cutoff for genomic follow-up analysis.
15. On this date, the typology presented by the Delta variant, depending on the level reported in the 2,814 cases as of September 3, 2021, this is divided as follows: a) 32.94% of the case studies have shown a mild level, which includes outpatient care cases, b) 35.54% corresponds to a moderate level, which includes the presence of both outpatient-asymptomatic and outpatient-symptomatic cases, and c) 31.52% have presented serious level where hospitalized patients and the re-registration of deaths associated with the presence of (SARS-CoV-2) COVID-19 are identified.
16. In Graph 4 as a combined effect to present the impact of the fatality rate and positivity index for the records of (SARS-CoV-2) COVID-19 in the indigenous language-speaking population, with accumulated figures as of 3 September 2021, a spatial ordering of the federative entities is observed where the State of Puebla presents the highest fatality rate (30.5%) and a high index of positivity (53.3%). While the states of Hidalgo, Veracruz de Ignacio de la Llave and Oaxaca, are located as the states with a high index of positivity and fatality rate.
17. From the information presented in Figure 5 above, it is highlighted that the five states that registered high rates of severity in infection in patients who speak an indigenous language by (SARS-CoV-2) COVID-19, at the national level and with Accumulated figure as of September 3, 2021, are the State of Puebla with 52.2%, the State of Mexico with 46.4%, the State of Chiapas with 46.2%, the State of Veracruz de Ignacio de la Llave with 45.2% and the State of Campeche with 43.9%.

18. The identification of the states that present a high fatality rate and positivity index for the records of (SARS-CoV-2) COVID-19 in the indigenous language-speaking population, and that identify the states of Puebla, Hidalgo, Veracruz de Ignacio de la Llave and Oaxaca, in these points the presence of the variant has been registered $VOC_{i\,j}^{Delta}$ = *[Delta (WHO label) ], [G/478K.V1 (GISAID clade) ],*[B.1.617.2+AY.x (Pango lineage)], with a registry of 308 cases of the 2,814 registered at the national level, being the State of Veracruz de Ignacio de la Llave that concentrates 153 identified cases, followed by the State of Hidalgo with 100 cases, the State of Puebla with 35 cases and the State of of Oaxaca with 20 cases, with figures as of September 3, 2021.
19. Of the 308 cases identified in the indicated states, the age group that concentrates the largest number of cases of the variant $VOC_{i\,j}^{Delta}$ = *[Delta (WHO label) ], [G/478K.V1 (GISAID clade) ],*[B.1.617.2+AY.x (Pango lineage)], They are between the ages of 21 and 40, with a record of 141 cases; followed by the interval from 41 years to 60 years with 93 cases identified. It is important to highlight that of the 308 cases identified in the selected states, 155 cases are women and 153 cases are men. See Table 13 at the end of the document.

The presence of (SARS-CoV-2) COVID-19 in the population that indicated that they speak an indigenous language shows that 36.25% of the accumulated cases registered in the Epidemiological Surveillance System for Viral Respiratory Diseases (SVEERV), as of September 3, 2021, registered a positive result for (SARSCoV-2) COVID-19, with the presence of positive results being greater in men with 53.43% while women only represented 46.57%.

In 10 states, 72.67% of the indigenous language-speaking population is concentrated, with a positive result for (SARS-CoV-2) COVID-19 and who have received hospital care, highlighting the State of Yucatan, which concentrates 18.51%. of patients, followed by the State of Oaxaca with 12.60%, the State of Mexico with 9.02%, the State of Puebla with 8.30%, the State of Hidalgo with 7.65%, the State of Quintana Roo with 5.34%, the State of of Guerrero with 5.10%, the State of Veracruz with 4.85% and the State of Michoacán de Ocampo with 3.57%.

As an indicator of the hospital medical care strategy, considering the level of symptoms and severity present in indigenous-language speaking patients, only 30.68% received hospital medical care. It is important to highlight that men are the largest number of patients treated in this modality, representing 58.43% while women correspond to 41.57%.

**Discussion**

The presence of (SARS-CoV-2) COVID-19 in Mexico, taking into account the age, sex and comorbidities present in indigenous-language speaking patients who have received a hospital medical care strategy, and depending on the level of severity who present symptoms, have had to be cared for in Intensive Care Units and, where appropriate, receive respiratory care and support through intubation techniques and procedures; It should be noted that only 777 indigenous language-speaking patients nationwide have received this type of specialized care, being 61.90% men and 38.10% women.

The number of accumulated deaths, at the national level, as of September 3, 2021 in the (SVEERV), correspond to 3,321 patients speaking an indigenous language who presented a positive record for (SARS-CoV-2) COVID-19; Of these, only 358 correspond to those who required admission to Intensive Care Units (ICU) and Intubation, that is, only 10.78%. As can be seen from Figure 3, section “c”, the comorbidities identified in the 358 death records do not highlight the presence of diabetes, hypertension or obesity, as aggravating factors of (SARS-CoV-2) COVID-19 in the population speaker of indigenous language in Mexico. On the contrary, the presence of smoking, pneumonia, chronic renal failure and COPD is observed with a higher prevalence in ages over 40 years.

This situation invites us to reflect on the impact that the (SARS-CoV-2) COVID-19 in the localities that are located in these states, since the presence of hospital care centers Covid 19, are mainly found in the cities that have a greater infrastructure for regional care, as is the case of urban centers, mainly. And on the other hand, it is important to note that in Mexico the presence of variants of (SARS-CoV-2) COVID-19 has been registered, the variant being $VOC_{i\,j}^{Delta}$ = *[Delta (WHO label) ], [G/478K.V1 (GISAID clade) ],*[B.1.617.2+AY.x (Pango lineage)], which is present in most of the genomic monitoring records, according to the information from the Information System of the Global Initiative on Sharing All Influenza Data (GISAID), as of September 3, 2021.

It is important to note that the 308 cases identified from the $VOC_{i\,j}^{Delta}$ , the State of Veracruz concentrates 49.68% of the cases identified in the selected states (Puebla, Hidalgo, Veracruz de Ignacio de la Llave and Oaxaca) which present a high fatality rate and a high index of positivity.

Faced with the onslaught caused by SARS-CoV-2) COVID-19 in the national territory, a vaccination strategy was launched, serving vulnerable groups, classifying by stages and scheduling for established age groups. It is important to note that, with information as of September 3, 2021, of the 308 study cases indicated in the selected federative entities, there is also a record of only 18 cases that received a vaccine, the following distribution being: a) State of the Hidalgo, 4 cases (1 Aztraseneca, 2 Cansino, 1 Pfizer); b) State of Veracruz de Ignacio de la Llave 13 cases (4 Aztraseneca, 1 Cansino, 8 Pfizer) and c) State of Puebla 1 (Unidentified).

The fatality rate as an indicator that shows the percentage of people killed by (SARSCoV-2) COVID-19 with respect to the total number of patients with a positive result, the State of Puebla presents the highest fatality rate for the indigenous language-speaking population in national level, being 30.5%, followed by the State of Campeche with a fatality rate of 25.1%, the State of Oaxaca with a fatality rate of 20.9%, d) the State of Quintana Roo with a rate of 20.9%, the State of Tlaxcala with a rate of 20.0% and the State of Veracruz de Ignacio de la Llave with an estimated fatality rate of 19.8%. Regarding the estimation of the positivity index in the indigenous language speaking population, at the national level with accumulated figures as of September 3, 2021, the State of Hidalgo presents the highest positivity index at the national level with 57.6% , followed by the State of Colima with 56.0%, the State of Veracruz de Ignacio de la Llave with 55.8%, the State of Oaxaca with 54.8%, the State of Puebla with 53.3% and the State of Quintana Roo with 52.9%. (See. Graph 3, presented previously).

It is highlighted that the population that speaks an indigenous language, at the national level as of September 3, 2021, the number of cases that have been treated in Intensive Care Units and that required breathing support through intubation are relatively few, that is, only 358. Why is the number of people served for this sector of the population significantly lower? An approximation to the answer may be that the registry of indigenous language speaking patients, contained in the Epidemiological Surveillance System for Viral Respiratory Diseases (SVEERV), requires precisely that, to be cared for in a hospital unit for the care of (SARS -CoV-2) COVID-19, and as has been indicated, the location of the care centers, the distance from the localities with an indigenous speaking population and the speed with which the transfer is required, are factors that limit access to health care services, and on the other hand, make more evident the lack of medical infrastructure near their population centers, thus increasing the inequality and marginalization suffered by these population centers.

And it is here, where in the presence of (SARS-CoV-2) COVID-19 the time, attention and access to medical services and hospital care is vital, because depending on age, sex and present comorbidities the rate of severity of the infection becomes more relevant and requires urgent attention.

Bibliography

1. Alvarez-Luna, R. et al. (2021). Marco de trabajo para la publicación de Datos Abiertos en Cuba. En: Revista Cubana de transformación digital. ISSN 2708-3411. Vol. 2, número 1, enero-marzo 2021. págs. 144-158. Recuperado de: https://rctd.uic.cu/rctd/article/view/109/43
2. Centers for Disease Control and Prevention. (2021). Emerging SARS-CoV-2 Variants. Enero 15 2021. Recuperado de:

https://www.cdc.gov/coronavirus/2019-ncov/more/science-and-research/scientific-brief-emerging-variants.html
3. Centros para el control y la prevención de enfermedades. (2021). Clasificaciones y definiciones de las variantes del SARS-CoV-2. Recuperado de: https://espanol.cdc.gov/coronavirus/2019-ncov/variants/variant-info.html#Interest
4. Demsar J, Curk T et al. (2013) Orange: Data Mining Toolbox in Python. En: Journal of Machine Learning Research 14 (Aug): 2349−2353. https://dl.acm.org/doi/pdf/10.5555/2567709.2567736
5. Global Initiative on Sharing All Influenza Data (GISAID) (2021). Open access to epidemic and pandemic virus data. Recuperado de: https://www.epicov.org/epi3/frontend#140535
6. Global Initiative on Sharing All Influenza Data (GISAID) (2021). Genomic epidemiology of hCoV-19. Recuperado de: https://www.gisaid.org/
7. Gobierno de México. Secretaría de Salud. (2021). Lineamiento estandarizado para la vigilancia epidemiológica y por laboratorio de la enfermedad respiratoria viral. Recuperado de: https://coronavirus.gob.mx/wp-content/uploads/2021/02/Lineamiento_VE_y_Lab_Enf_Viral_Ene-2021_290121.pdf
8. Gobierno de México. Secretaria de Salud. Bases de datos Covid-19 México. https://datos.gob.mx/busca/dataset/informacion-referente-a-casos-covid-19-en-mexico
9. INEGI, (2010). Catálogo de metadatos geográficos. Comisión Nacional para el Conocimiento y Uso de la Biodiversidad. División política municipal, 1:250000. 2010. Palabras clave: 2010, Área Geoestadística, División, Límite, Municipios. Fecha de publicación: 04-07-2011, del metadato 20-07-2017. Available at : http://www.conabio.gob.mx/informacion/gis/
10. Institute of Social and Preventive Medicine University of Bern, Switzerland& SIB Swiss Insitute of Bioinformatics, Switzerland (2021). Recuperado de: https://covariants.org/per-variant
11. Marie-Odile Safon y Véronique Suhard (2020) “Covid-19 Éléments de littérature scientifique Bibliographie thématique Septembre 2020. Centre de documentation del L'Institut de recherche et documentation en économie de la santé (Irdes), France.” ISBN 978-2-87812-526-9
12. Medel-Ramírez C. y Medel-López H. (2021). El (SARS-CoV-2) COVID 19 y su presencia en las zonas de hablantes de lengua indígenas en el Estado de Veracruz. Inédito.
13. Medel-Ramírez, Carlos and Medel-López, Hilario (2020): Data mining for the study of the Epidemic (SARS-CoV-2) COVID-19: Algorithm for the identification of patients speaking the native language in the Totonacapan area – Mexico. Avilable at https://www.researchgate.net/publication/343546537_Impact_of_SARS-CoV2_COVID_19_on_the_five_main_indigenous_languagespeaking_areas_in_Veracruz_Mexico_The_case_of_the_Totonacapan_area
14. Medel-Ramírez, Carlos and Medel-López, Hilario (2020): Impact of (SARS-CoV-2) COVID 19 on the five main indigenous language-speaking areas in Veracruz Mexico: The case of the Huasteco from theTantoyuca Area –

Mexico. Avilable at https://www.researchgate.net/publication/343570073_Impact_of_SARS-CoV2_COVID_19_on_the_five_main_indigenous_languagespeaking_areas_in_Veracruz_Mexico_The_case_of_the_Huasteco_from_the_Tantoyuca_Area

15. Medel-Ramírez, Carlos and Medel-López, Hilario (2020): Impact of (SARS-CoV-2) COVID 19 on the five main indigenous language- speaking areas in Veracruz Mexico: The case of the Nahuatl from the Pajapan Zone. – Mexico. Avilable at https://www.researchgate.net/publication/343798420_Impact_of_SARSCoV-2_COVID_19_on_the_five_main_indigenous_language_speaking_areas_in_Veracruz_Mexico_The_case_of_the_Nahuatl_from_the_Pajapan_Zone
16. Medel-Ramírez, Carlos and Medel-López, Hilario (2020): Impact of (SARS-CoV-2) COVID 19 on the five main indigenous language-speaking areas in Veracruz Mexico: The case of the Nahuatl of the Zongolica Zone – Mexico. Avilable at https://www.researchgate.net/publication/343760400_Impact_of_SARS-CoV2_COVID_19_on_the_five_main_indigenous_languagespeaking_areas_in_Veracruz_Mexico_The_case_of_the_Nahuatl_of_the_Zongolica_Zone
17. Medel-Ramírez, Carlos and Medel-López, Hilario (2020): Impact of (SARS-CoV-2) COVID 19 on the five main indigenous language-speaking areas in Veracruz Mexico: The case of the Otomi of the Ixhuatlan de Madero area – Mexico. Avilable at https://www.researchgate.net/publication/343628250_Impact_of_SARS-CoV2_COVID_19_on_the_five_main_indigenous_languagespeaking_areas_in_Veracruz_Mexico_The_case_of_the_Otomi_of_the_Ixhuatlan_de_Madero_area
18. Medel-Ramírez, Carlos and Medel-Lopez, Hilario, Data Mining for the Study of the Epidemic (SARS- CoV2) COVID-19: Algorithm for the Identification of Patients (SARS-CoV-2) COVID 19 in Mexico (June 3, 2020). Available at SSRN: https://ssrn.com/abstract=3619549 http://dx.doi.org/10.2139/ssrn.3619549
19. Ministerio de Salud de Argentina (2021). COVID 19 Situación actual de las nuevas variantes SARS-CoV-2. Informe técnico. Julio 2021. p. 2 Recuperado de: https://www.argentina.gob.ar/sites/default/files/2021/07/vigilancia-genomica-se26.pdf
20. Muik Alexander et al. (2021) Neutralization of SARS-CoV-2 lineage B.1.1.7 pseudovirus by BNT162b2 vaccine–elicited human sera. Science. 2021. Vol. 371, no.6534, p. 1152 Recuperado de: http://science.sciencemag.org/content/371/6534/1152.abstract
21. Junejo Y., *et al.* (2020) “Novel SARS-CoV-2/COVID 19: Origin, pathogenesis, genes and genetic variations, immune responses and phylogenetic analysis”, Gene Reports. Vol. 20, 2020, 100752, p. 2
22. Reuters. (2020).¿Cuáles son y en qué fase están las posibles vacunas contra la COVID 19. Recuperado de: https://www.eitb.eus/es/noticias/sociedad/detalle/7622841/listado-posibles-vacunas-covid19-noviembre-2020/

23. Secretaría de Salud (2021). Información referente a casos COVID-19 en México. Recuperado de: https://datos.gob.mx/busca/dataset/informacion-referente-a-casos-covid-19-en-mexico
24. Software Orange Data Mining version 3.26.1 https://orange.biolab.si
25. Tyrrel, D. A. J., J. D. Almedia, D. M. Berry, C. H. Cunningham, D. Hamre, M. S. Hofstad, L. Malluci, and K. McIntosh. (1968). Coronavirus. Nature 220:650. Recuperado de: https://www.ncbi.nlm.nih.gov/pmc/articles/PMC7182643/
26. Weiner L.P. (1987) Coronaviruses: A Historical Perspective. In: Lai M.M.C., Stohlman S.A. (eds) Coronaviruses. Advances in Experimental Medicine and Biology, vol 218 p. 2. Springer, Boston, MA. https://doi.org/10.1007/978-1-4684-1280-2_1
27. World Health Organization. (2020). Laboratory testing for coronavirus disease (COVID-19) in suspected human cases. Interim guidance. 19 March 2020. Recuperado de: https://apps.who.int/iris/handle/10665/331501?locale-attribute=es&
28. World Health Organization. (2021) Tracking SARS-CoV-2 variants. Recuperado de: https://www.who.int/en/activities/tracking-SARS-CoV-2-variants/
29. World Health Organization (2021). Genomic sequencing of SARS-CoV-2. A guide to implementation for maximum impact on public health. Recuperado de: https://www.who.int/publications/i/item/9789240018440

a. Reference to a dataset:

30. Raw data can be retrieved from the Github repository https://github.com/CMedelR/dataCovid19/

# Annexes

Table 1
Indigenous language speaking patients at the national level, according to result for (SARS-CoV-2) COVID-19 until September 3, 2021

| Clasificación | Femenino | Masculino | Total |
|---|---|---|---|
| Caso de COVID-19 confirmado por asociación clínica epidemiológica | 570 | 589 | 1,159 |
| Caso de COVID-19 confirmado por comité de dictaminación | 29 | 62 | 91 |
| Caso de SARS-COV-2 confirmado | 9,316 | 10,728 | 20,044 |
| Inválido por laboratorio | 34 | 38 | 72 |
| No realizado por laboratorio | 260 | 246 | 506 |
| Caso sospechoso | 2,746 | 2,602 | 5,348 |
| Negativo a SARS-COV-2 | 17,067 | 14,452 | 31,519 |
| Total | 30,022 | 28,717 | 58,739 |

Source: Own elaboration with information as of September 3, 2021

Table 2
Indigenous language-speaking patients nationwide, according to a positive result for (SARS-CoV-2) COVID-19 as of September 3, 2021

| Clasificación | Femenino | Masculino | Total |
|---|---|---|---|
| Caso de COVID-19 confirmado por asociación clínica epidemiológica | 570 | 589 | 1,159 |
| Caso de COVID-19 confirmado por comité de dictaminación | 29 | 62 | 91 |
| Caso de SARS-COV-2 confirmado | 9,316 | 10,728 | 20,044 |
| Total | 9,915 | 11,379 | 21,294 |

Source: Own elaboration with information as of September 3, 2021

Table 3
Indigenous language-speaking patients nationwide with a positive result for (SARS-CoV-2) COVID-19 according to the Medical Treatment Strategy until September 3, 2021

| | Estrategia de Tratamiento Médico | | |
|---|---|---|---|
| Clasificación | Ambulatorio | Hospitalario | Total |
| Femenino | 7,199 | 2,716 | 9,915 |
| Masculino | 7,562 | 3,817 | 11,379 |
| Total | 14,761 | 6,533 | 21,294 |

Source: Own elaboration with information as of September 3, 2021

Table 4
Percentage distribution of indigenous language speaking patients with a positive result for (SARS-CoV-2) COVID-19 according to the Medical Treatment Strategy and state as of September 3, 2021

| | Estrategia de Tratamiento Médico | | | |
|---|---|---|---|---|
| Entidad Federativa | Ambulatorio | Hospitalario | Total | % |
| Yucatán | 2,242 | 1,209 | 3,451 | 18.51% |
| Oaxaca | 1,577 | 823 | 2,400 | 12.60% |
| México | 727 | 589 | 1,316 | 9.02% |
| Puebla | 427 | 542 | 969 | 8.30% |
| Hidalgo | 759 | 500 | 1,259 | 7.65% |
| Quintana Roo | 627 | 349 | 976 | 5.34% |
| Guerrero | 1,221 | 333 | 1,554 | 5.10% |
| Ciudad de México | 2,000 | 317 | 2,317 | 4.85% |
| Veracruz de Ignacio de la Llave | 380 | 317 | 697 | 4.85% |
| Michoacán de Ocampo | 302 | 233 | 535 | 3.57% |
| Jalisco | 193 | 136 | 329 | 2.08% |
| Chihuahua | 236 | 135 | 371 | 2.07% |
| Sonora | 278 | 130 | 408 | 1.99% |
| Chiapas | 148 | 126 | 274 | 1.93% |
| Campeche | 163 | 120 | 283 | 1.84% |
| San Luis Potosí | 1,518 | 115 | 1,633 | 1.76% |
| Nayarit | 150 | 93 | 243 | 1.42% |
| Baja California | 220 | 89 | 309 | 1.36% |
| Guanajuato | 166 | 54 | 220 | 0.83% |
| Morelos | 79 | 46 | 125 | 0.70% |
| Tabasco | 345 | 45 | 390 | 0.69% |
| Sinaloa | 87 | 42 | 129 | 0.64% |
| Nuevo león | 188 | 41 | 229 | 0.63% |
| Tlaxcala | 50 | 30 | 80 | 0.46% |
| Durango | 101 | 29 | 130 | 0.44% |
| Querétaro | 180 | 19 | 199 | 0.29% |
| Zacatecas | 64 | 17 | 81 | 0.26% |
| Colima | 41 | 15 | 56 | 0.23% |
| Tamaulipas | 86 | 14 | 100 | 0.21% |
| Coahuila de Zaragoza | 74 | 10 | 84 | 0.15% |
| Aguascalientes | 15 | 9 | 24 | 0.14% |
| Baja California Sur | 117 | 6 | 123 | 0.09% |
| TOTAL | 14,761 | 6,533 | 21,294 | 100.00% |

Source: Own elaboration with information as of September 3, 2021

Table 5
Indigenous-language speaking patients with a positive result for (SARS-CoV-2) COVID-19 that required intubation, according to sex as of September 3, 2021

| Requirieron intubación | Mujeres | Hombres | Total |
|---|---|---|---|
| Si | 296 | 481 | 777 |
| No | 2,394 | 3,293 | 5,687 |
| No aplica | 7,199 | 7,562 | 14,761 |
| Se ignora | 26 | 43 | 69 |
| Total | 9,915 | 11,379 | 21,294 |

Source: Own elaboration with information as of September 3, 2021

Table 6
Deaths of patients who speak an indigenous language with a positive result for (SARS-CoV-2) COVID-19, by sex as of September 3, 2021

| Clasificación | Mujeres | Hombres | Total |
|---|---|---|---|
| Caso de COVID-19 confirmado por asociación clínica epidemiológica | 101 | 149 | 250 |
| Caso de COVID-19 confirmado por comité de dictaminación | 29 | 62 | 91 |
| Caso de SARS-COV-2 confirmado | 1,118 | 1,862 | 2,980 |
| Total | 1,248 | 2,073 | 3,321 |

Source: Own elaboration with information as of September 3, 2021

Table 7
Deaths of patients who speak an indigenous language with a positive result for (SARS-CoV-2) COVID-19 that required attention in the Intensive Care Unit, according to sex as of September 3, 2021

| | Estrategia de Tratamiento Médico en UCI | | |
|---|---|---|---|
| Requirieron Unidad de Cuidados Intensivos | Mujeres | Hombres | Total |
| Si | 122 | 236 | 358 |
| No | 984 | 1,570 | 2,554 |
| No aplica | 130 | 234 | 364 |
| Se ignora | 12 | 33 | 45 |
| Total | 1,248 | 2,073 | 3,321 |

Fuente: Elaboración propia con información al 3 de septiembre de 2021

# Table 8

VOC ALPHA

| Linage | Clade G | Clade GR | Clade GRY | Total |
|---|---|---|---|---|
| B.1.1.7 | 4 | 76 | 864 | 944 |
| Q.1 | 0 | 0 | 3 | 3 |
| Q.3 | 0 | 1 | 15 | 16 |
| Total | 4 | 77 | 882 | 963 |

Fuente: Elaboración propia.

VOC BETA

| Lineage | Clade GH | Total |
|---|---|---|
| B.1.351 | | 11 |
| Total | 11 | 11 |

Fuente: Elaboración propia.

VOC GAMMA

| Linage | Clade G | Clade GK | Clade GR | Total |
|---|---|---|---|---|
| P.1 | 6 | 0 | 1,413 | 1,419 |
| P.1.1 | 0 | 0 | 2 | 2 |
| P.1.10 | 0 | 1 | 7 | 8 |
| P.1.10.2 | 0 | 0 | 17 | 17 |
| P.1.11 | 1 | 0 | 1 | 2 |
| P.1.2 | 0 | 0 | 20 | 20 |
| P.1.7 | 0 | 0 | 3 | 3 |
| P.1.8 | 0 | 0 | 1 | 1 |
| P.1.9 | 0 | 0 | 1 | 1 |
| Total | 7 | 1 | 1,465 | 1,473 |

Fuente: Elaboración propia.

VOC DELTA

| Linage | Clade G | Clade GK | Clade GV | Clade O | Total |
|---|---|---|---|---|---|
| AY.10 | 1 | 0 | 0 | 0 | 1 |
| AY.12 | 2 | 23 | 0 | 0 | 25 |
| AY.13 | 1 | 10 | 0 | 0 | 11 |
| AY.15 | 0 | 5 | 0 | 0 | 5 |
| AY.19 | 0 | 3 | 0 | 0 | 3 |
| AY.2 | 2 | 1 | 0 | 0 | 3 |
| AY.20 | 70 | 1,232 | 0 | 0 | 1,302 |
| AY.21 | 1 | 3 | 0 | 0 | 4 |
| AY.24 | 0 | 1 | 0 | 0 | 1 |
| AY.25 | 0 | 11 | 0 | 0 | 11 |
| AY.3 | 5 | 125 | 0 | 0 | 130 |
| AY.4 | 0 | 7 | 0 | 0 | 7 |
| AY.5 | 0 | 1 | 0 | 0 | 1 |
| AY.9 | 0 | 1 | 0 | 0 | 1 |
| B.1.617.2 | 94 | 1,213 | 1 | 1 | 1,309 |
| Total | 176 | 2,636 | 1 | 1 | 2,814 |

Fuente: Elaboración propia.

VOI ETA

No presente en México

VOI IOTA

| Linage | Clade G | Clade GH | Total |
|---|---|---|---|
| B.1.526 | 2 | 27 | 29 |
| Total | 2 | 27 | 29 |

Fuente: Elaboración propia.

VOI KAPPA

| Linage | Clade G | Total |
|---|---|---|
| B.1.617.1 | | 2 |
| Total | 2 | 2 |

Fuente: Elaboración propia.

VOI LAMBDA

| Linage | Clade G | Clade GR | Total |
|---|---|---|---|
| C.37 | 1 | 54 | 55 |
| Total | 1 | 54 | 55 |

Fuente: Elaboración propia.

VOI MU

| Linage | Clade GH | Total |
|---|---|---|
| B.1.621 | | 144 |
| B.1.621.1 | | 4 |
| Total | 148 | 148 |

Fuente: Elaboración propia.

Source: Own elaboration with information from the Information System of the Global Initiative on Sharing All Influenza Data (GISAID), as of September 3, 2021

Table 9
Genomic surveillance of the VOC variant i j Delta = [Delta (WHO label)], [G / 478K.V1 (GISAID clade)], [B.1.617.2 + AY.x (Pango lineage)] in Mexico until September 3, 2021

| Estatus del paciente | Clade G | Clade GK | Clade GV | Clade O | Total |
|---|---|---|---|---|---|
| Liberado | 10 | 905 | 1 | 1 | 917 |
| Vivir | - | 9 | - | - | 9 |
| Atención ambulatoria en vivo | - | 1 | - | - | 1 |
| Ambulatorio | 112 | 859 | - | - | 971 |
| Moderar | - | 4 | - | - | 4 |
| Asintomático - Ambulatorio | - | 5 | - | - | 5 |
| Ambulatorio asintomático | - | 1 | - | - | 1 |
| Asintomático y ambulatorio | - | 2 | - | - | 2 |
| Ambulatorio sintomático | - | 1 | - | - | 1 |
| Sintomático | - | 1 | - | - | 1 |
| Sintomático - Ambulatorio | - | 10 | - | - | 10 |
| Sintomático y ambulatorio | - | 5 | - | - | 5 |
| Hospitalizado | 53 | 793 | - | - | 846 |
| Fallecido | 1 | 33 | - | - | 34 |
| Fatal | - | 7 | - | - | 7 |
| Total | 176 | 2,636 | 1 | 1 | 2,814 |

Source: Own elaboration with information from the Information System of the Global Initiative on Sharing All Influenza Data (GISAID), as of September 3, 2021.

Table 10
Number of cases identified by type of CLADE
in the identified variant [Delta (WHO label)], [G / 478K.V1 (GISAID clade)], [B.1.617.2 + AY.x (Pango lineage)] in selected states as of September 3, 2021

| Clade | Clade G | Clade GK | Total |
|---|---|---|---|
| Estado de Puebla | 1 | 34 | 35 |
| Estado de Hidalgo | 5 | 95 | 100 |
| Estado de Veracruz | 11 | 142 | 153 |
| Estado de Oaxaca | 5 | 15 | 20 |
| Total | 22 | 286 | 308 |

Source: Own elaboration with information from the Information System of the Global Initiative on Sharing All Influenza Data (GISAID), as of September 3, 2021.

Table 11

Number of registered cases of the variant [Delta (WHO label)], [G / 478K.V1 (GISAID clade)], [B.1.617.2 + AY.x (Pango lineage)] in selected states as of September 3 from 2021

| Entidad Federativa | Mujeres | Hombres | Total |
|---|---|---|---|
| Estado de Puebla | 16 | 19 | 35 |
| Estado de Hidalgo | 46 | 54 | 100 |
| Estado de Veracruz | 82 | 71 | 153 |
| Estado de Oaxaca | 11 | 9 | 20 |
| Total | 155 | 153 | 308 |

Source: Own elaboration with information from the Information System of the Global Initiative on Sharing All Influenza Data (GISAID), as of September 3, 2021.

Table 12

Type of vaccine applied in registered cases with an identified variant [Delta (WHO label)], [G / 478K.V1 (GISAID clade)], [B.1.617.2 + AY.x (Pango lineage)] in selected states as of September 3, 2021

| Entidad Federativa | Aztraseneca | Cancino | Pfizer | Ninguna | Total |
|---|---|---|---|---|---|
| Estado de Puebla | - | - | - | 1 | 1 |
| Estado de Hidalgo | 1 | 2 | 1 | - | 4 |
| Estado de Veracruz | 4 | 1 | 8 | - | 13 |
| Estado de Oaxaca | - | - | - | - | - |
| Total | 5 | 3 | 9 | 1 | 18 |

Source: Own elaboration with information from the Information System of the Global Initiative on Sharing All Influenza Data (GISAID), as of September 3, 2021.

Table 13

Number of registered cases of the variant [Delta (WHO label)], [G / 478K.V1 (GISAID clade)], [B.1.617.2 + AY.x (Pango lineage)] in selected states by age groups and sex as of September 3, 2021

| Entidad Federativa | Mujeres | | | | Hombres | | | | Total | | | | |
|---|---|---|---|---|---|---|---|---|---|---|---|---|---|
| | De 0 a 20 años | De 21 a 40 años | De 41 a 60 años | De 60 años y más | De 0 a 20 años | De 21 a 40 años | De 41 a 60 años | De 60 años y más | De 0 a 20 años | De 21 a 40 años | De 41 a 60 años | De 60 años y más | Total |
| **Estado de Puebla** | 0 | 12 | 3 | 1 | 4 | 11 | 2 | 2 | 4 | 23 | 5 | 3 | 35 |
| **Estado de Hidalgo** | 3 | 19 | 15 | 9 | 6 | 23 | 16 | 9 | 9 | 42 | 31 | 18 | 100 |
| **Estado de Veracruz** | 4 | 36 | 31 | 11 | 14 | 30 | 19 | 8 | 18 | 66 | 50 | 19 | 153 |
| **Estado de Oaxaca** | 2 | 5 | 4 | 0 | 0 | 5 | 3 | 1 | 2 | 10 | 7 | 1 | 20 |
| **Total** | 9 | 72 | 53 | 21 | 24 | 69 | 40 | 20 | 33 | 141 | 93 | 41 | 308 |

Source: Own elaboration with information from the Information System of the Global Initiative on Sharing All Influenza Data (GISAID), as of September 3, 2021.